\documentclass[10pt]{article}
\usepackage{jheppub}
\usepackage{graphicx}
\usepackage{dcolumn}
\usepackage{bm}
\usepackage{amssymb}
\usepackage{amsmath}
\usepackage{epsfig}    
\usepackage{color}
\usepackage{slashed}
\usepackage{hhline}
\usepackage{xcolor}
\usepackage{cancel}
\usepackage{fmtcount} 
\usepackage{array,multirow}

\def\be{\begin{equation}}
\def\ee{\end{equation}}

\newcommand{\bea}{\begin{eqnarray}}
\newcommand{\eea}{\end{eqnarray}}

\newcommand{\wt}{\widetilde}

\newcommand{\la}{\lambda}
\newcommand{\tb}{\tan\beta}
\newcommand{\hpm}{H^\pm}

\title{Same sign trilepton as signature of charged Higgs in two Higgs doublet model}
\author[a]{Tanmoy Mondal,}
\author[b]{Prasenjit Sanyal}
\affiliation[a]{School of Physics, Korea Institute for Advanced Study, Seoul 02455, Republic of Korea}
\affiliation[b]{Asia Pacific Center for Theoretical Physics, Pohang 37673, Republic of Korea}
\emailAdd{tanmoy@kias.re.kr}
\emailAdd{prasenjit.sanyal@apctp.org}

\abstract{
We explored the prospect of looking for a fermiophobic charged Higgs ($\hpm$) via the same sign trilepton 
signal at the LHC. A fermiophobic scenario appears in the type-I two Higgs doublet model where the 
coupling of the $\hpm$ with the Standard Model fermions is inversely proportional to $\tb$. Almost all the experimental searches 
rely on the fermionic production and decay of the charged Higgs. Consequently, the limit on $\hpm$ for 
fermiophobic scenarios is non-existent unless $\tb$ is small. We show that for a fermiophobic case, the 
electroweak production of $\hpm$ is dominant for most of the parameter space. Subsequent bosonic decay 
of the charged and neutral Higgses give rise to the same sign trilepton signal. With a thorough phenomenological 
analysis, we demonstrate that the same sign trilepton signal can be an excellent complementary search 
to explore the high $\tb$ regions.
}

\preprint{KIAS-P21032, APCTP Pre2021-021}
\date{\today}
\begin{document}
\maketitle

\section{Introduction}
 Extension of the Higgs sector is ubiquitous in physics beyond the Standard Model (BSM), and two Higgs  Doublet Model (2HDM) is one of the simplest extensions containing two scalar doublets instead of one for electroweak symmetry breaking. To avoid the dangerous flavour changing neutral current, the Yukawa interactions are restricted\cite{Gunion:1989we,Djouadi:2005gj,Branco:2011iw} and the phenomenology of the BSM scalars vary based on the Yukawa structure. 
A charged Higgs boson ($\hpm$) would be one of the most striking signals of an extended Higgs sector 
like 2HDM. The Large Hadron Collider (LHC) has performed several searches for the charged Higgs. 
For most of the cases the collider studies look for $\hpm$ produced in association with a top quark and 
decays to jets or $\tau$ leptons~\cite{ATLAS:2013uxj,ATLAS:2014otc,CMS:2015lsf,CMS:2015yvc,ATLAS:2016avi,ATLAS:2018gfm,ATLAS:2018ntn,CMS:2018dzl,CMS:2019bfg,ATLAS:2020jqj,ATLAS:2021upq}. Production of a charged Higgs via vector 
boson fusion is also explored in~\cite{CMS:2017fgp,ATLAS:2018iui}. Discovery prospects of charged Higgs in model independent framework are studied in~\cite{Coleppa_2020,Coleppa:2021wjx}.

The LHC searches are motivated by the Yukawa structure of type-II 2HDM and supersymmetry. Thus these searches 
can not explore an extensive part of $\hpm$ phenomenology where the $\hpm$ is fermiophobic. The type-I  2HDM at 
the large $\tb$ is one of the simplest realizations of fermiophobic scalars. In this scenario, the typical 
production of charged Higgs via the top channel becomes very low. As a result, the existing limit on charged Higgs 
in the type-I 2HDM is essentially non-existent for $\tb$ larger than five~\cite{Chen:2019pkq,Kling:2020hmi}. 
Thus, to investigate the vast region with large $\tb$, we need to look for $\tb$ independent channels. 
Hence, exploration of bosonic decay of the BSM scalars is essential, which is governed by the gauge coupling. Another aspect that determines the signature 
of a $\hpm$ is the BSM scalar spectrum. If a BSM scalar lighter than $\hpm$ exist, then the signature of the 
$\hpm$ varies accordingly.


The importance of the bosonic decay model of the $\hpm$ has been identified before. The decay of
 $\hpm$ to $W^\pm$ and a neutral scalar has been studied for both type-II and type-I 2HDM model with $\hpm tb$ 
 associated production of charged Higgs~\cite{Coleppa:2014cca,Enberg:2014pua,Kling:2015uba,Akeroyd:2016ymd,Arhrib:2016wpw,Alves:2017snd,Arhrib:2019ywg,Arhrib:2020tqk,Sanyal:2019xcp} and in linear colliders~\cite{Akeroyd_1999,Demirci_2020}. 
 For fermiophobic cases, the $\hpm tb$ coupling is $\tb$ suppressed, and the channel becomes irrelevant for large $\tb$. On the other hand, the electroweak production of $\hpm$~\cite{Kanemura:2001hz,Cao:2003tr,Belyaev:2006rf,Chun:2018vsn,Bahl:2021str} in association with a neutral (pseudo)scalar depends on the 
 gauge coupling and dominates over the top associated production at large $\tb$. 
In type-I 2HDM,  the electroweak production of a light charged Higgs can give rise to multi-photon or 
multi-boson final state~\cite{Arhrib:2017wmo,Enberg:2018pye,Enberg:2018nfv,Arhrib:2021xmc,Wang:2021pxc,Akeroyd:2003bt,Akeroyd_2004}. 
However, if the BSM scalars are heavier than the observed Higgs boson, they dominantly 
decay to a pair of massive gauge bosons, and these analyses can not be applied. Hence 
it is crucial to consider how to look for a $\hpm$ if it is heavier than the SM Higgs boson to understand the experimental discovery potential of a charged Higgs.

Our goal is to present a complementary search strategy for the charged and neutral Higgses, which 
are heavier than the SM Higgs boson and display fermiophobic nature. We consider 
electroweak production of the charged Higgs and its subsequent decay to the heavy CP even neutral scalar 
$\hpm\to W^\pm H$. We found that this channel remains dominant for a wide range of parameter space and 
give rise to a distinctive $5W$ final state. Although it is possible to explore the $5W$ final state via 
trilepton signature, the signature will be overwhelmed by the huge background coming from $WZ, t\bar t$ and $t\bar t V$ channels. Going further to four lepton final state decreases the signal cross-section significantly, and 
it will compete with another substantial $ZZ/Z\gamma$ background.

In this work, we have shown that the most promising signal is the same sign trilepton (SS3L) final state. Background for this 
process is rare. We have done a realistic analysis of the SS3L final state for the fermiophobic BSM sector by using 
the type-I 2HDM as a proxy scenario. We show that the proposed signal will discover or rule out a significant 
parameter space for BSM scalars even at an integrated luminosity of 300 $fb^{-1}$. The exclusion bounds can 
be as low as $\tb=2$, and any $\tb$ larger than that will be covered by the high luminosity LHC with an 
integrated luminosity of 3000 $fb^{-1}$. The proposed signal complement the existing search strategies to 
expand the reach of LHC searches for charged Higgs. 
To show robustness of our study, we have studied the scenarios for three different mass gaps between charged Higgs and the heavy CP even neutral scalar. For a mass gap of 
120 GeV, all relevant processes apart from $H\to W^+W^-$ is on-shell and when mass gap is 85 GeV, the process
$A\to ZH$ becomes off-shell. It is possible to accommodate even lower mass gap of 60 GeV where all the processes 
including $\hpm\to W^\pm H$  become off-shell.

The paper is organized as follows: In Sec.~\ref{sec:model} we briefly discuss the type-I 2HDM model, 
including the existing theoretical and experimental bounds on the parameter space. Then we motivate towards 
the possible SS3L signature in Sec.~\ref{sec:ss3l} where we also provide the details of the collider analysis. 
The results are presented in Sec.~\ref{sec:result} and we conclude in Sec.~\ref{sec:conclusion}.

\section{The Model and Experimental Bounds}\label{sec:model}
Here we give a brief overview of the type-I 2HDM and discuss possible phenomenology and experimental constraints. 

\subsection{The 2HDM Type-I Model}\label{subsec:model}
The 2HDM model consists of two scalar doublets $\Phi_1$ and $\Phi_2$ with hypercharge 
$Y=1$~\cite{Gunion:1989we,Djouadi:2005gj,Branco:2011iw}. Flavor changing neutral current (FCNC) interaction appears at 
tree level when both the doublets couple to the fermions. It is possible to suppress FCNC at tree 
level by imposing an additional $\mathbb{Z}_2$ symmetry such that $\Phi_1\rightarrow -\Phi_1$ and 
$\Phi_2\rightarrow \Phi_2$. The fermions are also charged appropriately under the discrete symmetry.
The $\mathbb{Z}_2$ symmetric scalar potential is,
\begin{eqnarray}
\nonumber V_{\mathrm{2HDM}} &=& -m_{11}^2\Phi_1^{\dagger}\Phi_1 - m_{22}^2\Phi_2^{\dagger}\Phi_2 -\Big[m_{12}^2\Phi_1^{\dagger}\Phi_2 + \mathrm{h.c.}\Big]
+\frac{1}{2}\lambda_1\left(\Phi_1^\dagger\Phi_1\right)^2+\frac{1}{2}\lambda_2\left(\Phi_2^\dagger\Phi_2\right)^2 \\
&& +\lambda_3\left(\Phi_1^\dagger\Phi_1\right)\left(\Phi_2^\dagger\Phi_2\right)+\lambda_4\left(\Phi_1^\dagger\Phi_2\right)\left(\Phi_2^\dagger\Phi_1\right)
+\Big\{ \frac{1}{2}\lambda_5\left(\Phi_1^\dagger\Phi_2\right)^2+  \rm{h.c.}\Big\}.
\label{eq:2hdm-pot}
\end{eqnarray}

The $\mathbb{Z}_2$ symmetry is softly broken by the dimensionful coupling $m_{12}^2$ and we have considered the parameters $m_{12}^2$ and $\lambda_5$ to be real assuming CP invariance. We parameterize the doublets in the following way,
\be 
\Phi_j=\begin{pmatrix}
        H_j^+\\\dfrac{1}{\sqrt2}(v_j + h_j + i A_j) 
       \end{pmatrix},\hspace{1cm} j = 1,2.
\ee
The scalar spectrum consists of five massive states, two CP-even neutral scalars $h$ and $H$, one CP-odd pseudoscalar $A$, and a pair of charged Higgs $H^{\pm}$. The mass eigenstates can be expressed 
in terms of the gauge eigenstates:
\begin{align}\label{2hdm_scalar_basis}
 \begin{pmatrix} H  \\ h \end{pmatrix} =  \begin{pmatrix}  c_{\alpha} && s_{\alpha} \\ -s_{\alpha} &&  c_{\alpha} \end{pmatrix}
  \begin{pmatrix} h_1  \\ h_2 \end{pmatrix}, ~~~ A=-s_\beta \;A_1 + c_\beta \;A_2\quad\textrm{and} ~~~ H^{\pm}=-s_\beta\; H_1^{\pm} + c_\beta\; H^{\pm}_2.
 \end{align}
Here $s_\alpha = {\rm sin}~\alpha$, $c_\beta = {\rm cos}~ \beta$ etc. and 
${\rm tan}~\beta = \cfrac{v_2}{v_1}$ .
The CP-even state $h$ is identified as the SM-like Higgs boson with mass $m_h \approx 125$ GeV.

\begin{table}[t]
\begin{center}
\begin{tabular}{|c||c|c|c|c|c|c|c|c|c|}
\hline
2HDM& $\xi_h^u$ & $\xi_h^d$ & $\xi_h^\ell$
& $\xi_H^u$ & $\xi_H^d$ & $\xi_H^\ell$
& $\xi_A^u$ & $\xi_A^d$ & $\xi_A^\ell$ \\ \cline{2-10}

type-I& $c_\alpha/s_\beta$ & $c_\alpha/s_\beta$ & $c_\alpha/s_\beta$
& $s_\alpha/s_\beta$ & $s_\alpha/s_\beta$ & $s_\alpha/s_\beta$
& $\cot\beta$ & $-\cot\beta$ & $-\cot\beta$ \\
 \hline
\end{tabular}
\end{center}
 \caption{The Yukawa multiplicative factors in type I 2HDM}
\label{Tab:YukawaFactors}
\end{table}

Based on the $\mathbb{Z}_2$ charge assignment of the fermions, there are  four possible types of Yukawa 
structures, and in this article, we will consider the type-I 2HDM, where the fermions are even under 
$\mathbb{Z}_2$ symmetry and thus couple only with $\Phi_2$. The relevant Yukawa Lagrangian is given by,
\begin{equation}\label{eq:yukawa}
-{\cal L}_Y= Y^u\bar{ Q_L} \wt \Phi_2 u_R + Y^d  \bar{ Q_L} \Phi_2 d_R+Y^e\bar{ l_L} \Phi_2 e_R + h.c.,
\end{equation}
where $\wt \Phi_2=i\sigma_2\Phi_2^*$. After the symmetry breaking, we can write the Yukawa Lagrangian in 
terms of mass eigenstates,
\begin{eqnarray}
\nonumber \mathcal L_{\mathrm{Yuk,I}}^{\mathrm{Physical}} &=&
-\sum_{f=u,d,\ell} \frac{m_f}{v}\left(\xi_h^f\overline{f}hf +
\xi_H^f\overline{f}Hf - i\xi_A^f\overline{f}\gamma_5Af \right) \\
 &&-\left\{ \frac{\sqrt{2}V_{ud}}
{v}\overline{u}\left(\xi_A^{u} m_{u} P_L+\xi_A^{d} m_{d} P_R\right)H^{+}d  +
\frac{\sqrt{2}m_l}{v}\xi_A^l\overline{v}_LH^{+}l_R + \mathrm{h.c.}\right\}.
\label{eq:L2hdm}
\end{eqnarray}
Here the up-type quarks, down-type quarks, and charged leptons are denoted as $u$, $d$, and $l$ respectively. 
The Yukawa multiplicative factors ($\xi^f_\phi$) for the type-I scenario are given in Tab.~\ref{Tab:YukawaFactors}. 
Note that we can write couling of $H$ with fermions ($s_\alpha/s_\beta$) as $\cos(\beta-\alpha) -\sin(\beta-\alpha)/\tan\beta$.
Hence the BSM Higgs becomes fermiophobic when $\tan\beta = \dfrac{\sin(\beta-\alpha)}{\cos(\beta-\alpha)}$. 
We will use this relation later in our analysis.

For completeness, let us provide the relevant scalar couplings. 
In 2HDM, the  couplings of scalars with a pair of gauge bosons are given by
\cite{Gunion:1989we,Djouadi:2005gj}: 
\begin{equation}
g_{hVV}=\mathrm{sin}(\beta-\alpha)g_{hVV}^{\mathrm{SM}},\,\,\,\,g_{HVV}=\mathrm{cos}(\beta-\alpha)g_{hVV}^{\mathrm{SM}},\,\,\,\,g_{AVV}=0,
\end{equation}
where $V$ = $Z,\,W^\pm$. 
The couplings of $Z$ boson with the neutral scalars are,
\begin{align}
&hAZ_\mu:\,\frac{g_Z^{}}{2}\cos(\beta-\alpha)(p+p')_\mu,\quad
 HAZ_\mu:\,-\frac{g_Z^{}}{2}\sin(\beta-\alpha)(p+p')_\mu,
\label{hhV}
\end{align}
where $p_\mu$ and $p'_\mu$ are
outgoing four-momenta of the first and the second scalars, respectively,
and $g_Z^{}=g^{}/\cos\theta_W^{}$.
Couplings involving the charged scalar is given by,
\begin{align}
H^\pm hW^\mp_\mu:\,\mp i\frac{g}{2}\cos(\beta-\alpha)(p+p')_\mu,\quad
 H^\pm HW^\mp_\mu:\,\pm i\frac{g}{2}\sin(\beta-\alpha)(p+p')_\mu,\quad
H^\pm AW^\mp_\mu:\,\frac{g}{2}(p+p')_\mu
\label{hhV}
\end{align}
The trilinear scalar coupling which governs the decay of heavy Higgs to a pair of SM Higgs bosons is~\cite{PhysRevD.77.115013}
\be\label{eq:Hhh}
\lambda_{Hhh} = -\frac{\cos(\beta-\alpha)}{v~\sin(2\beta)^2}\left[\left(2m_h^2+m_H^2\right)\sin2\alpha\sin2\beta - \left(3\sin2\alpha-\sin2\beta\right)m_{12}^2 \right].\ee


\subsection{Theoretical Constraints}
The quartic couplings in Eq.~\ref{eq:2hdm-pot} are restricted by vacuum stability of the potential, tree unitarity and perturbativity.
The conditions are,
\begin{itemize}
 \item Perturbativity :  $|\la_i| < 4 \pi$
 \item Vacuum stability~\cite{Gunion:2002zf}:
\bea
0<\la_1,~\la_2 < 4\pi,  \ \la_3 > - \sqrt{\la_1 \la_2}, \ \ \& \ \ \la_3 + \la_4 - |\la_5| > - \sqrt{\la_1 \la_2} \ .
\eea
\item Unitarity constraints are discussed in ~\cite{Kanemura:1993hm,Akeroyd:2000wc}.
\end{itemize}
We have used the \texttt{2HDMC-1.8.0}~\cite{Eriksson:2009ws} package to check the above constraints before 
any phenomenological analysis.

The presence of two Higgs doublets modifies the electroweak oblique parameters~\cite{Peskin:1991sw}. The 
$T$-parameter restricts large mass splitting among the components of the doublet, and we have assumed that 
the pseudoscalar is degenerate with the charged Higgs ($m_A=m_{\hpm}$), thus evading the precision 
constraints.

\subsection{Experimental Constraints}
\begin{figure}[t!]
\begin{center}
\includegraphics[width=77mm]{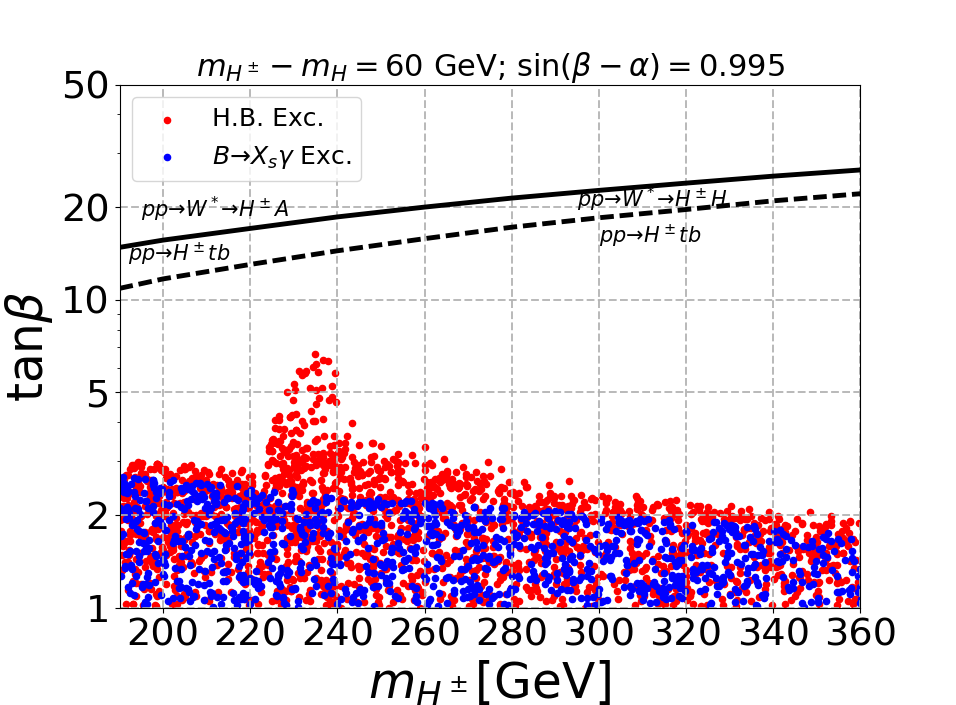}
\includegraphics[width=77mm]{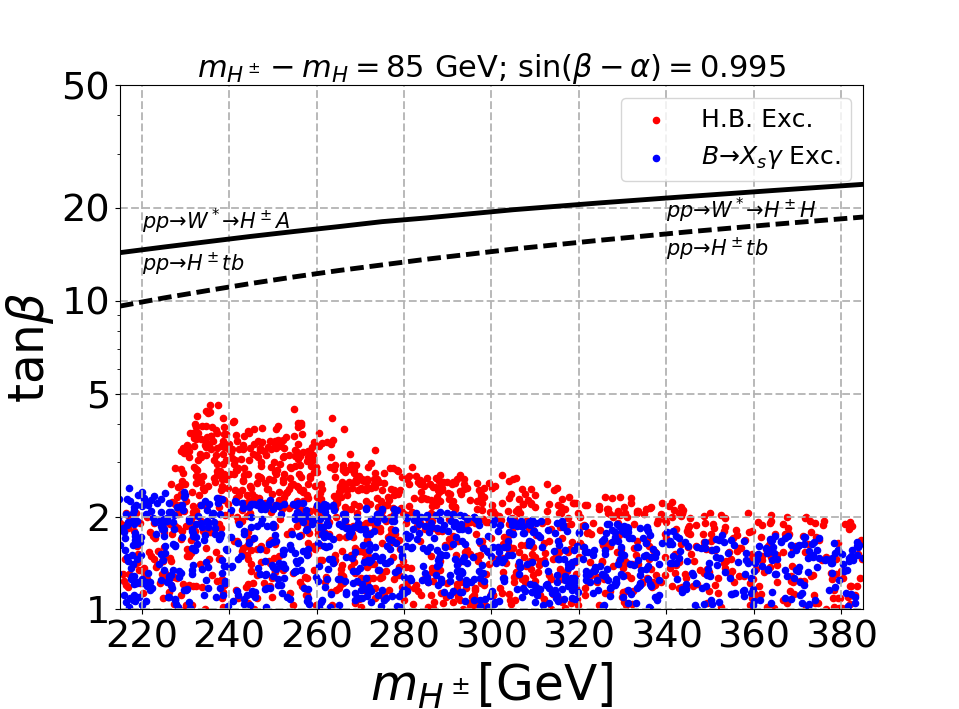}
\end{center}
\begin{center}
\includegraphics[width=77mm]{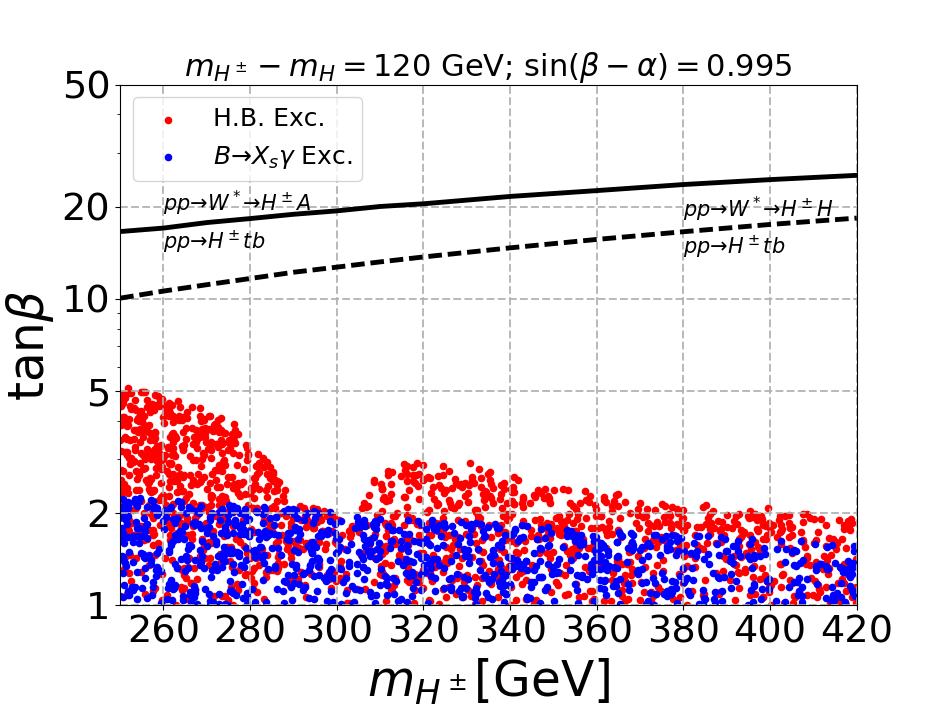} 
\end{center} 
\caption{Experimental constraints on $m_{H^\pm}-\tb$ plane is shown. The exclusion limits coming from LHC
are shown in red points, and the blue points show the excluded regions coming 
from the BR($B \to X_s \gamma$) constraint. Above the dashed(solid) black line the electroweak 
charged Higgs production $pp\to W^* \to \hpm H(pp\to W^*\to\hpm A)$ dominates over the top associated 
channel ($pp \to \hpm tb$).}
\label{fig:excluded region}
\end{figure}

To obtain the limits coming from experimental searches at the LHC, we have used the public code \texttt{HiggsBounds-5.10.2}~\cite{Bechtle:2013wla,Bechtle:2020pkv}.
We are interested in the scenario where the CP-even Higgs $H$ is lighter than $\hpm$ and $A$  which 
are degenerate. For our analysis we choose three mass differences, viz. 60 GeV, 85 GeV and 120 GeV. A wider mass gap 
makes $\hpm$ heavy and the production cross-section decreases. We choose the mixing angle 
$\sin(\beta - \alpha)=0.995$. We then scan the CP-even heavy Higgs in the range $m_H \in 
[130-300]$ GeV, $\tan\beta \in [1,50]$ and $m_{12}^2 \in [0,m_H^2 \sin\beta\cos\beta]$.
We used \texttt{HiggsSignals-2.6.2}~\cite{Bechtle_2014,Bechtle:2020uwn} to satisfy the SM Higgs signal strength measurements and 
since we are very close to the alignment limit which is $\sin(\beta - \alpha)=1$, the above mentioned parameter space easily satisfy the constraints on the SM Higgs boson measurements.

\begin{figure}[h!]
\begin{center}
\includegraphics[width=50mm]{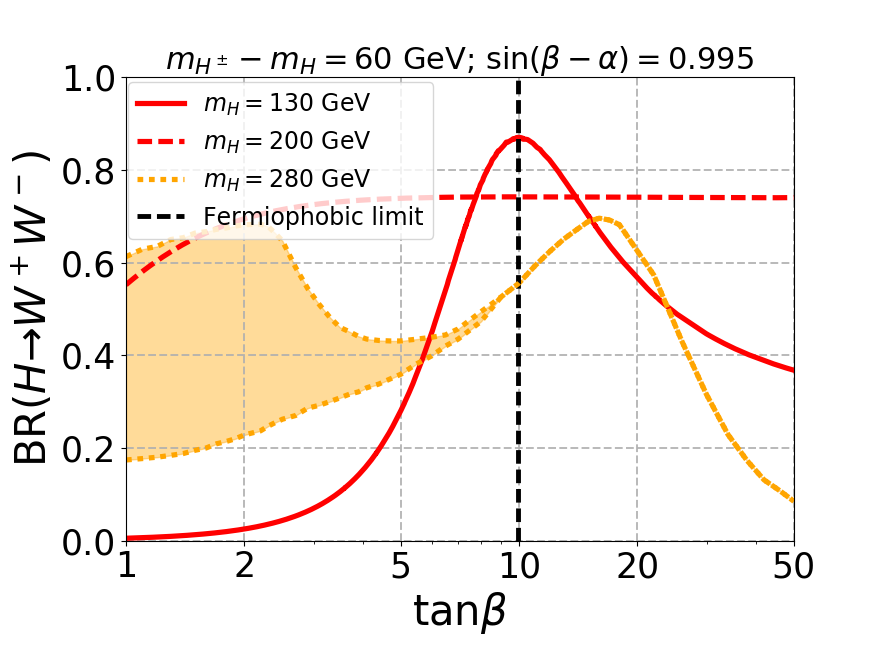}
\includegraphics[width=50mm]{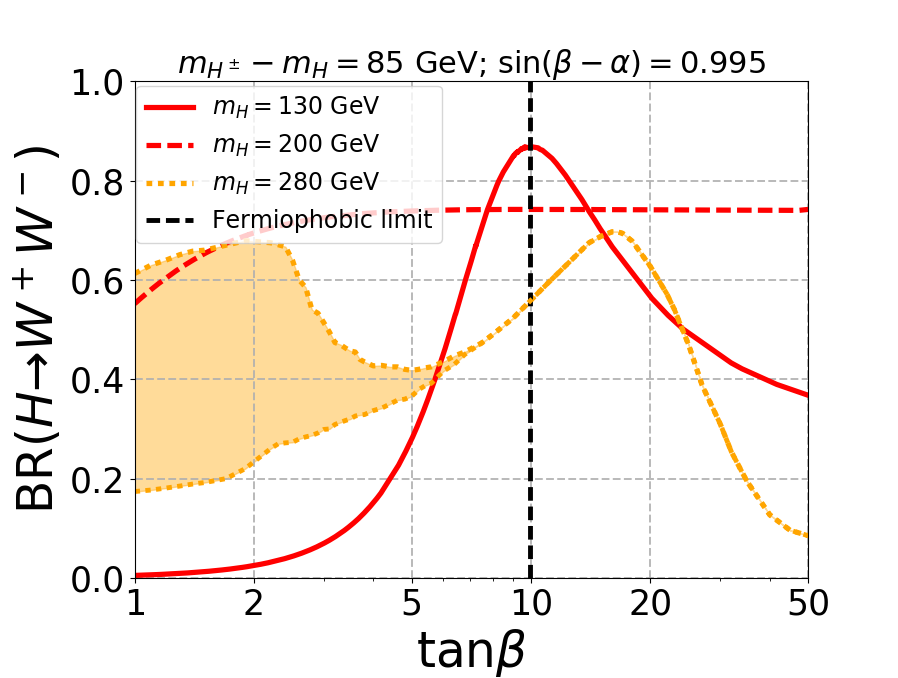}
\includegraphics[width=50mm]{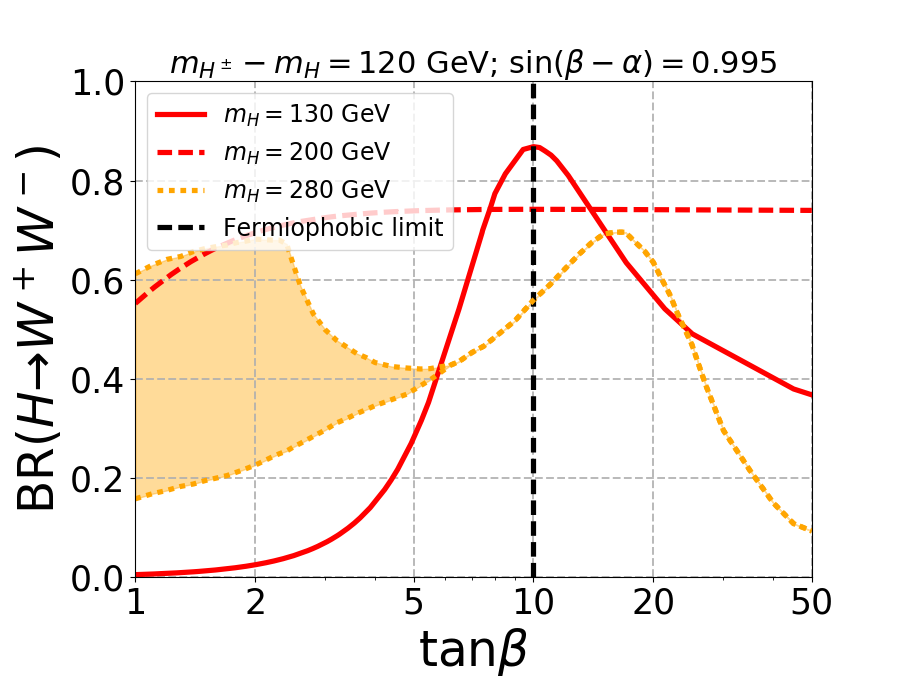} 
\includegraphics[width=50mm]{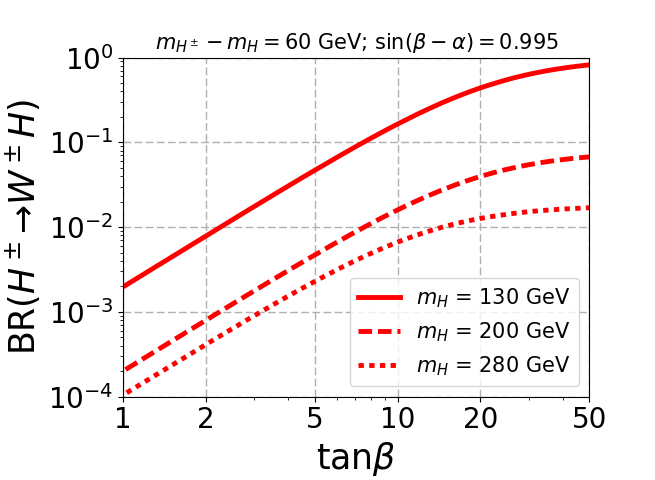}
\includegraphics[width=50mm]{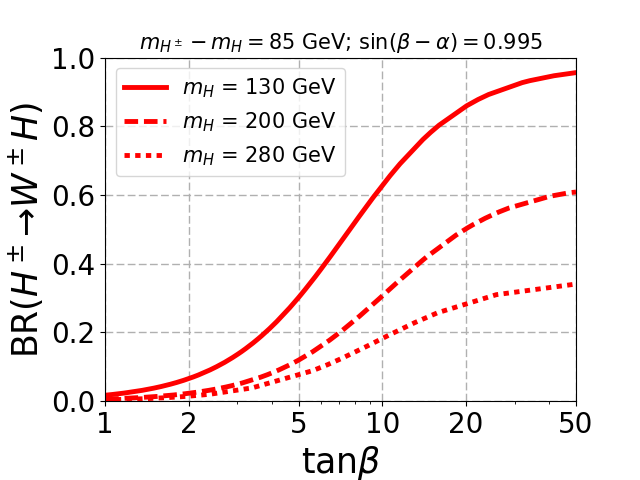}
\includegraphics[width=50mm]{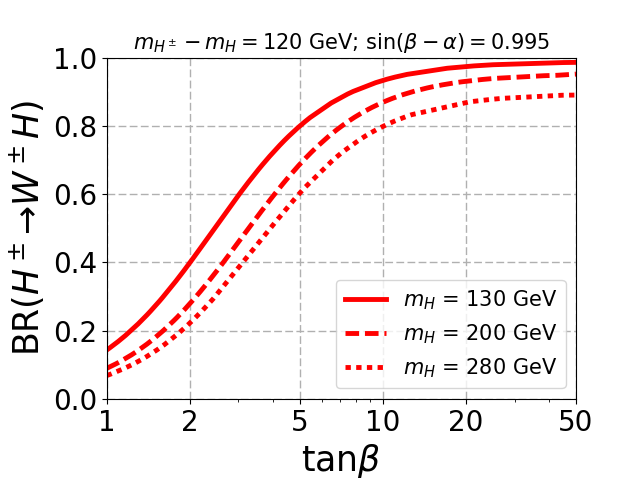}
\includegraphics[width=50mm]{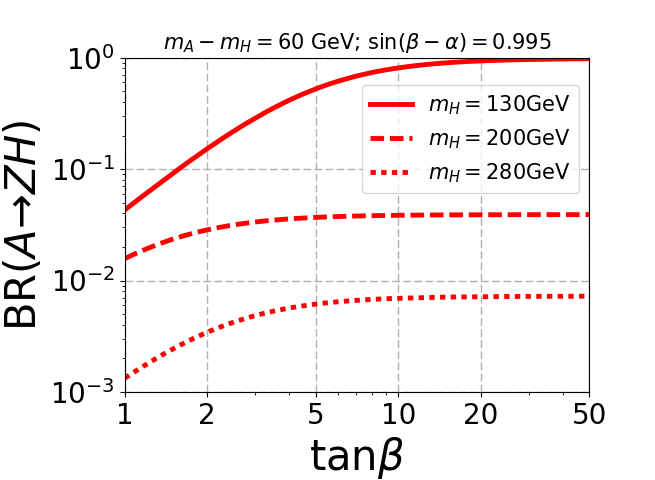}~
\includegraphics[width=50mm]{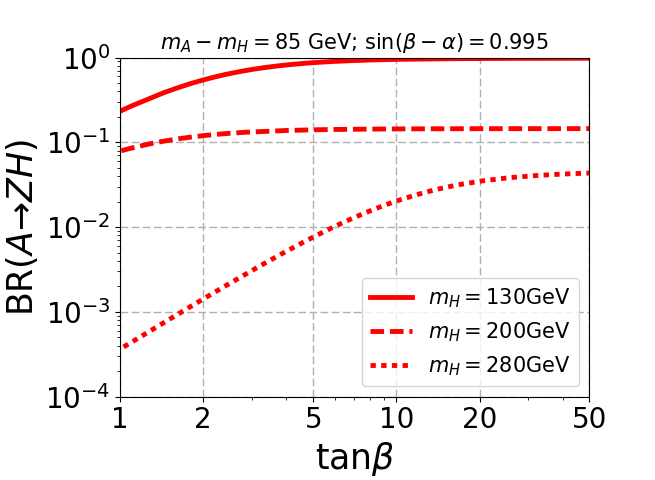}~
\includegraphics[width=50mm]{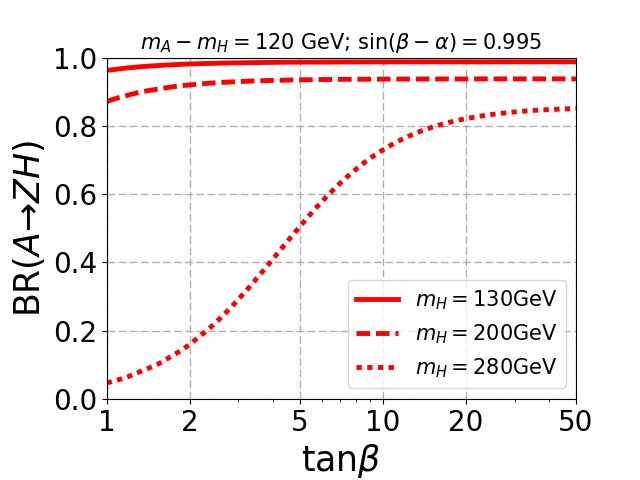}
\end{center}
\caption{Relevant branching ratio of the additional Higgs bosons decaying to gauge boson are shown for 
three different mass gaps. See text for the explanation.}
\label{fig:Branching ratios}
\end{figure}

In Fig.~\ref{fig:excluded region} we have shown exclusion in $m_{\hpm}-\tb$ plane for different mass gaps 
between $\hpm$ and $H$. The red points depict the limit coming from LHC measurements and we found that 
the constraint on the $m_{\hpm}-\tb$ plane is governed by the following LHC searches as obtained by \texttt{HiggsBounds-5.10.2}: 
$A\to Zh$~\cite{ATLAS:2015kpj,CMS:2018xvc}, $A\to ZH$~\cite{ATLAS:2018oht}, 
$A/H\to \tau\tau$~\cite{CMS:2015mca}, $H\to ZZ$~\cite{CMS:2017vpy} or $\gamma\gamma$~\cite{ATLAS:2014jdv}, 
$\hpm\to tb$~\cite{ATLAS:2018ntn} 
and also the SM Higgs boson decay to 4-leptons~\cite{CMS:2013wyb}. 
In the 2HDM scenario the observation of $B\to X_s \gamma$~\cite{HFLAV:2016hnz,Misiak:2017bgg} constrain the charged 
Higgs mass and the limit is shown in blue points using the package \texttt{SUPERISO-2.5}~\cite{Mahmoudi_2009}. For the type-I 2HDM scenario, the production of neutral Higgses 
($H/A$) via gluon fusion as well as the production of charged Higgs via the top associated channel ($ pp\to \hpm t b$) 
is suppressed by $(\tb)^2$. Hence, the limit coming from present LHC searches (red points) primarily restricts 
the small $\tb$ region, and the limits will remain weak even in HL-LHC~\cite{Chen:2019pkq}. 
We would like to point out that the limit coming from $pp\to A\to Z h$ is the strongest when all 
the BSM scalars are degenerate~\cite{Kling:2020hmi}. However, in our case, the decay
of the pseudoscalar to $Z H$ channel is substantial, and consequently, the experimental sensitivity is poor. 
Similarly, the $B\to X_s\gamma$ exclusion is weak due to the fermiophobic nature of fermionic coupling of $\hpm$.
In the top left (mass gap of 60 GeV) of Fig.~\ref{fig:excluded region} when the decay $A\to Z h $ opens up, the exlusion limit becomes 
stronger which is clearly visible when $m_A(=m_\hpm)$ crosses 220 GeV. The same argument applies to the top right panel (mass gap of 85 GeV). 
In the bottom panel (mass gap of 120 GeV) $A\to Z h$ is always open, but the limit becomes weak as $m_\hpm$ increases. The limit again 
becomes strong when $H\to ZZ$ opens up and this explains the dip in the exclusion region obtained from \texttt{HiggsBounds-5.10.2}.

%
From Fig.~\ref{fig:excluded region} it is evident that the existing LHC search strategies are inadequate, and we 
need to explore possible signatures which will be effective in the high $\tb$ regime. As $\tb$ increases, the
production of $\hpm$ in association with top quark decreases but the electroweak(EW) production 
is independent of $\tb$. The black solid and dashed lines in Fig.~\ref{fig:excluded region} depict 
when the EW process dominates over top quark processes, and it is evident that we need to consider the EW process for large $\tb$. In the next
section, we have explored the phenomenology of $\hpm$ produced alongside $H$ and $A$. We will show that the 
exploration via gauge boson channels can significantly enhance the reach of LHC for the $\hpm$ search.

\section{Same sign trilepton as probe of charged Higgs}\label{sec:ss3l}

In the previous section, we demonstrated that the existing LHC limit on the BSM scalars in the type-I 2HDM is rather weak. We can 
improve it by exploring the EW production of $\hpm$ and its decay into gauge bosons. Here we will show that the most 
promising channel is the same sign trilepton (SS3L) final state. Such signal originates via the following processes\footnote{Another possible source of SS3L is through the charged Higgs pair creation mode: $pp \to H^+ H^- \to (W^+ H)(W^- H) \to (W^+ W^+ W^-)(W^- W^+ W^-) \to  3\ell^\pm  \cancel{\it{E}}_{T}+X$. However, the charged Higgs pair production via $Z^*/\gamma$ channel is much smaller than the charged current channel and hence does not affect our result.}:
\begin{eqnarray}\label{eq:signal}
p p \rightarrow  W^{* \pm} \rightarrow \hpm H \rightarrow (W^\pm H) (W^+ W^-) \rightarrow (W^\pm W^+ W^-)(W^+ W^-) &\rightarrow& 3\ell^\pm  \cancel{\it{E}}_{T}+X\nonumber\\
p p \rightarrow  W^{* \pm}  \rightarrow \hpm A \rightarrow (W^\pm H) (Z H) \rightarrow (W^\pm W^+ W^-)(Z W^+ W^-) &\rightarrow& 3\ell^\pm  \cancel{\it{E}}_{T}+X.
\end{eqnarray} 
Here $X$ is any additional jets and/or leptons.
Since the pseudoscalar $A$ is degenerate with $\hpm$, the bottom process will be subdominant compared to 
the $\hpm H$ channel. Before going into the phenomenological study of the signal, let us first discuss the decay of the BSM Higgs to the gauge bosons. We compute the decay widths and the branching ratios using \texttt{2HDMC-1.8.0}.

\subsection{Bosonic decays of the Higgs bosons}

In Fig.~\ref{fig:Branching ratios} the branching ratios of additional Higgs bosons are shown for different 
mass differences, where we fix $\sin(\beta-\alpha)=0.995$. The top panel shows the decay of $H$ to $W^+W^-$ final 
state for different $m_H$ as a function of $\tb$. When $m_H$ is smaller than $2 m_W$ (solid red curve), the $W^+W^-$ 
branching ratio peaks only at the fermiophobic limit (indicated by the vertical black dashed line), which occurs 
for $\tb=10$ when $\sin(\beta-\alpha)=0.995$. As we move away from the fermiophobic limit, the branching ratio
decreases slightly but remains large enough to be explored at the LHC. As $m_H$ increases and on-shell decay to $W$ 
pair opens up, $H\to W^+W^-$ becomes the dominant decay mode irrespective of $\tb$ as shown by the red-dashed curves 
in the top panel. When $m_H > 2m_h$ the di-Higgs channel opens up, which decreases the $H\to W^+W^-$ branching ratio. 
The $H-h-h$ coupling as shown in Eq.~\ref{eq:Hhh} depends on $m_{12}^2$, and for small $\tb$, the allowed range 
of $m_{12}^2$ varies substantially, which changes the $H\to h h$ decay width. Consequently, the BR($H\to W^+W^-$) 
also varies, which is shown by the yellow region. As $\tb$ increases, the value of $m_{12}^2$ approaches 
$m_H^2\sin\beta\cos\beta$ to satisfy the stability and perturbativity constraints, and the yellow regions become 
a single curve. When $\tb$ becomes very large, the $H-h-h$ coupling dominates, which decrease the $W^+W^-$ branching 
substantially. The figures in the top planel look independent of mass gap since the mass gap changes the masses 
of $\hpm$ (and $A$) which affects only the di-photon branching ratio, a loop suppressed process.

The middle panel in Fig.~\ref{fig:Branching ratios} shows the branching ratio of charged Higgs to $HW^\pm$ channel. 
As $\tb$ increases, the decay to $tb$ channel decreases, and BR($\hpm\to HW^\pm$) as well as BR($\hpm\to hW^\pm$) 
increases. For the mass gap of 60 GeV and 85 GeV, the kinematic suppression is strong enough to decrease BR($\hpm\to HW^\pm$) 
as $m_{H}$ (and therefore $m_{H^\pm}$) increases. When the mass gap increases (120 GeV), the kinematic suppression becomes irrelevant, and 
BR($\hpm\to HW^\pm$) becomes dominant since decay to $H^\pm \to hW^\pm$ is $\cos(\beta-\alpha)$ suppressed.

The lower panel in Fig.~\ref{fig:Branching ratios} depicts the decay of pseudoscalar to $ZH$ channel. When 
the mass gap is smaller than the required for on-shell production of $ZH$, the decay $A\to hZ$ dominates despite 
$\cos(\beta-\alpha)$ suppression. As the mass gap increases, $A$ mostly decays to $ZH$, as shown in the right plot of 
the bottom panel. When $A\to t\bar t$ opens up, the $ZH$ branching ratio decreases substantially for low $\tb$, 
which is evident in the red dotted curves.

From Fig.~\ref{fig:Branching ratios} it is evident that branching ratio of the bosonic decay channels $\hpm(A)\to HW^\pm(Z)$ with respect to the mass of $H$ (and therefore $H^\pm$)   
decreases as the mass gap decreases. In addition the selection efficiency of the decay products coming from the off-shell decay of $\hpm$ and $A$ will be
very low as we decrease the mass gap. Thus, the signal process is not suitable for very small mass gap where both the dominant and the subdominant channels are mediated through off-shell decays of $H^\pm$ and $A$. Hence, we have not considered $m_\hpm-m_H < 60$ GeV despite the large production cross-section of $\hpm$. Note that, for small mass gap the decay of $\hpm(A) \to W^\pm(Z)h$ becomes dominant and might be useful.

\subsection{Event Generation and Signal Selection }

\begin{table}[!t]
\begin{center}
\begin{tabular}{|c|c|c|c|c|c|c|c|}
\hline
\multicolumn{8}{|c|}{Signal cross-sections at $\sqrt{s} = 13$ TeV}  \\
\hline
\hline
\multicolumn{2}{|c|}{Selection cuts}  
& MG5 & SS3L & $p_{T}(\ell)$ \& $\cancel{\it{E}}_{T}$ & $\Delta R_{ll}$ \& $\Delta R_{lj}$ & $Z$-veto & b-veto \\
\hline
\multirow{2}{*}{$\Delta m = 60$ GeV} & Signal 1 [fb] & 0.741 & 0.00237 
  & 0.00113 & 0.000975 & 0.000962 & 0.0007\\ 
\cline{2-8}
 & Signal 2 [fb] & 0.0537 & 0.000224 & 0.0000698 & 0.0000604 & 0.0000590 & 0.0000406 \\
\cline{1-8}
\multirow{2}{*}{$\Delta m = 85$ GeV} & Signal 1 [fb] & 9.18 & 0.0372 
  & 0.0205 & 0.0186 & 0.0182 & 0.0131\\ 
\cline{2-8}
 & Signal 2 [fb] & 1.2 & 0.00754 & 0.0048 & 0.00418 & 0.00399 & 0.00276 \\
\cline{1-8}
\multirow{2}{*}{$\Delta m = 120$ GeV} & Signal 1 [fb] & 15.84 & 0.0668 
  & 0.0405 & 0.0370 & 0.0364 & 0.0260\\ 
\cline{2-8}
 & Signal 2 [fb] & 6.13 & 0.0374 & 0.0252 & 0.0229 & 0.0190 & 0.0113 \\
\cline{1-8}
\end{tabular}
\caption{Effects of the selection cuts on the signal cross-sections at $\sqrt{s} = 13$ TeV LHC for the mass differences $\Delta m = m_{H^\pm}-m_{H} = 60$ GeV, 85 GeV and 120 GeV. Signal 1 and Signal 2 are the dominant ($H^\pm H$ channel) and the subdominant ($H^\pm A$ channel) signals respectively. We fixed the parameters $m_{H} = 175$ GeV, $\tan\beta = 10$, $\sin(\beta - \alpha)=0.995$, $m_A = m_{H^\pm}$ and $m_{12}^2$ is properly taken to satisfy the theoretical constraints. Note that the cross-sections at the MG5 level (first column) is up to the 5W production which further decays into SS3L.}
\label{tab:sig}
\end{center}
\end{table}

As discussed at the beginning of the section, the charged Higgs will be produced via the following process: 
$pp\to W^{\pm*} \to \hpm H/A$ and the subsequent bosonic decay of the Higgses yield the same sign 
trilepton signal ($3\ell^\pm+ \cancel{\it{E}}_{T}$) in association with  additional jets and/or leptons. We applied uniform K-factor of 1.35 \cite{Bahl:2021str} for the signal cross-section. 
For event generation, the type-I 2HDM model is implemented in \texttt{FeynRules-2.3}\cite{Alloul:2013bka} 
and both the signal and the background events are generated using 
\texttt{MadGraph5-aMC@NLO-2.6.6}\cite{Alwall:2011uj,Alwall:2014hca}. 
We used \texttt{PYTHIA-8.2} \cite{Sjostrand:2006za,Sjostrand:2014zea} for showering and hadronization, 
and for detector simulation we used \texttt{Delphes-3.4.2}\cite{deFavereau:2013fsa} using  anti-kt algorithm \cite{Cacciari:2008gp} for jet clustering with radius parameter $R = 0.5$ and $p_T(j) > 20$ GeV.
In \texttt{Delphes} we have used the following lepton identification and isolation criteria:
\begin{itemize}
 \item Electrons and muons should have $p_T(\ell) > 10$ GeV and the electron efficiency
 is taken as 95\% and 90\% for $|\eta| < 1.5$ and $1.5 < |\eta| < 2.5$ respectively. Efficiency for muon is
 98\% when $|\eta| < 1.5$ and 95\% if $1.5 < |\eta| < 2.4$.
 \item The lepton isolation ($R^\ell_{Iso}$) is defined as the ratio of the sum of $p_T$ of all 
 objects within a cone with $\Delta R=0.2$ to the $p_T$ of the lepton. We demand that $R^\ell_{Iso} < 0.2$.
\end{itemize}
 {\bf Selection criteria:} A signal event is selected based on the following selection criteria:
\begin{itemize}
\item SS3L: We select events with three isolated leading leptons ($e,\mu$) with same sign. 
\item Lepton $p_T$ cuts: We impose the transverse momentum on the SS3L signal as, $p_{T}(\ell_1) > 30$ 
GeV, $p_{T}(\ell_2) > 30$ GeV and $p_{T}(\ell_3) > 20$ respectively where leptons are ordered according to
the transverse momentum.
\item Missing energy cut: We impose a nominal MET cut $\cancel{\it{E}}_{T} > 30$ GeV to reduce events with 
jets as fake leptons. 
\item Lepton and jet separation cuts: We impose the lepton-lepton separation, $\Delta R_{\ell\ell} > 0.4$ and 
lepton-jet separation cuts, $\Delta R_{\ell j} > 0.4$  where $\Delta R=\sqrt{\Delta\eta^2 + \Delta\phi^2}$.
\item $Z$-veto: If additional leptons with opposite sign to the tagged three same sign leptons are present 
in an event, we veto such events if any opposite-sign same flavor lepton pair combination satisfy the invariant mass condition
$80<m_{\ell^+\ell^-}<100$ GeV.  
\item $b$-veto: We veto events if there are any tagged $b$-jets. This minimizes the background comes from 
$t\bar t$ processes.
\end{itemize}

In Tab.~\ref{tab:sig} we have shown the cut flow for both the signals given in Eq.~\ref{eq:signal} where Signal 1 (Signal 2) refers to the $\hpm H~(\hpm A)$ production channels. For benchmark points we fixed  $m_{H} = 175$ GeV, $\tan\beta = 10$ and $\sin(\beta - \alpha)=0.995$ with three mass gaps $\Delta m =m_{H^\pm}-m_H = 60$ GeV, 85 GeV and 120 GeV.
As the mass gap increases, the cross-section increases due to the increase of BR($\hpm\to W^\pm H$) and BR($A \to Z H$). For the mass gap of 60 GeV large kinematic suppression comes in both the signals $\hpm H~(\hpm A)$ and for the mass gap of 85 GeV the kinematic suppression comes mostly to the subdominant ($\hpm A$) signal.


\begin{table}
\centering
\begin{tabular}{|p{3.3cm}|p{1.5cm}|p{1.5cm}|p{2cm}|p{2.1cm}|p{1.5cm}|p{1.5cm}|}
\hline
 \multicolumn{7}{|c|}{Background cross-sections at $\sqrt{s} = 13$ TeV} \\
 \hline
   \hline
   Backgrounds   & MG5 & SS3L & $p_T(\ell)$ \& $\cancel{\it{E}}_{T}$ & $\Delta R_{\ell\ell}$ \& $\Delta R_{\ell j}$ & $Z$-veto & $b$-veto \\
   \hline
   $WZ+$jets [fb] & 1360.80 & 0.0543 & 0.0122 & 0.0073 & 0.0065 & 0.0061 \\
   \hline
   $Z\ell^+\ell^-+$jets [fb] & 246.55 & 0.00991 & 0.00122 & 0.00083 & 0.00065 & 0.00061 \\
   \hline
    $ZZW+$jets [fb]  & 0.781 & 0.00881 & 0.00555 & 0.00525 & 0.001667 & 0.00155 \\
   \hline
   $h(>ZZ)W+$jets [fb] & 0.218 & 0.00210 & 0.000783 & 0.000616 & 0.000432 & 0.00041  \\
   \hline
   $h(>WW)t\bar{t}+$jets [fb] & 20.51 & 0.00842 & 0.00095 & 0.000875 & 0.000875 & 0.000123  \\  
   \hline
   $t\bar{t}W+$jets [fb] & 62.57 & 0.0878 & 0.0118 & 0.0103 & 0.0103 & 0.0018 \\
   \hline
   $t\bar{t}Z+$jets [fb] & 92.08 & 0.0321 & 0.00705 & 0.00648 & 0.00583 & 0.00092 \\
   \hline
\end{tabular}
\caption{Effects of the selection cuts on the cross-sections for the dominant backgrounds of SS3L at $\sqrt{s} = 13$ TeV LHC.}
\label{tab:BG}
\end{table}

\subsection{Background Estimation}
The SM backgrounds can be divided into irreducible and reducible backgrounds. The relevant irreducible backgrounds are $ZZW$ + jets (K-factor = 1.85 \cite{Alwall:2014hca}), $h(\to ZZ)W$ + jets (K-factor = 1.18 \cite{Alwall:2014hca}), $ZZZ$ + jets (K-factor = 1.31 \cite{Alwall:2014hca}), $h(\to ZZ)Z$ + jets (K-factor = 1.18 \cite{Alwall:2014hca}) and $h(\to ZZ)t\bar{t}$ + jets (K-factor = 1.17 \cite{LHCHiggsCrossSectionWorkingGroup:2016ypw}). The relevant reducible backgrounds are $WZ$ + jets (K-factor = 1.3 \cite{Campanario:2010hp}), $Z \ell^+ \ell^-$ + jets (K-factor = 1.7 \cite{Cascioli:2014yka}), $h(\to WW)t\bar{t}$ + jets (K-factor = 1.17 \cite{LHCHiggsCrossSectionWorkingGroup:2016ypw}), $t\bar{t}W$ + jets (K-factor = 1.22 \cite{Maltoni:2015ena}) and $t\bar{t}Z$ + jets (K-factor = 1.44 \cite{Maltoni:2015ena}). We generated all 
the background events matched up to one parton using the MLM scheme \cite{Artoisenet_2013}. The cross-sections and the cut flow for the dominant backgrounds at $\sqrt{s} = 13$ TeV according to the above mentioned selection 
cuts are given in Tab.~\ref{tab:BG}. The subdominant backgrounds contribute only at the order of  $10^{-4}$ $fb$. The total background cross-section at $\sqrt{s} = 13$ TeV is 0.01161 $fb$. For both the signal and the backgrounds we have included the decay of $W$ boson to $\tau$ lepton, which can further decay to light leptons and contribute to signal and background processes.

\begin{figure}[t]
\begin{center}
\includegraphics[width=80mm]{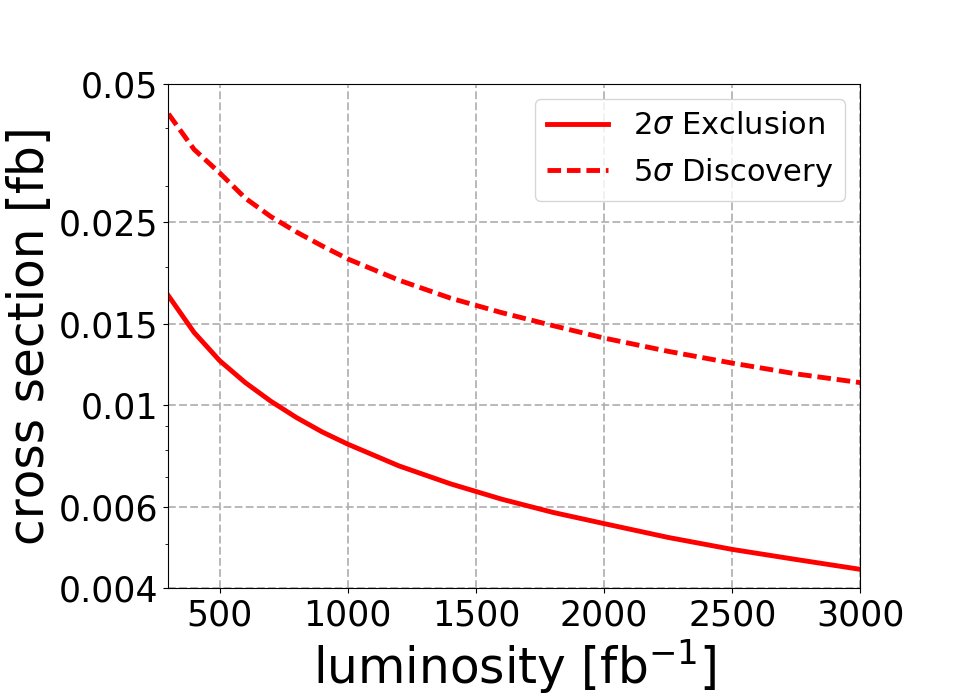}
\caption{The minimum signal cross-sections required for $2\sigma$ exclusion and $5\sigma$ discovery for luminosity ranging from $300$ $fb^{-1}$ and $3000$ $fb^{-1}$ at $\sqrt{s}=13$ TeV LHC.}
\label{fig:signal-x-section-lumi}
\end{center}
\end{figure}

\section{Results and Discussion}\label{sec:result}

\begin{figure}[t!]
\begin{center}
\includegraphics[width=77mm]{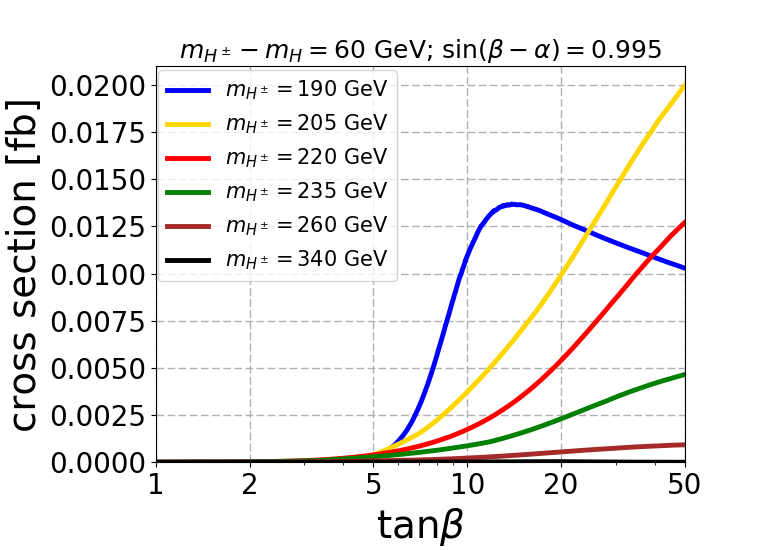}
\includegraphics[width=77mm]{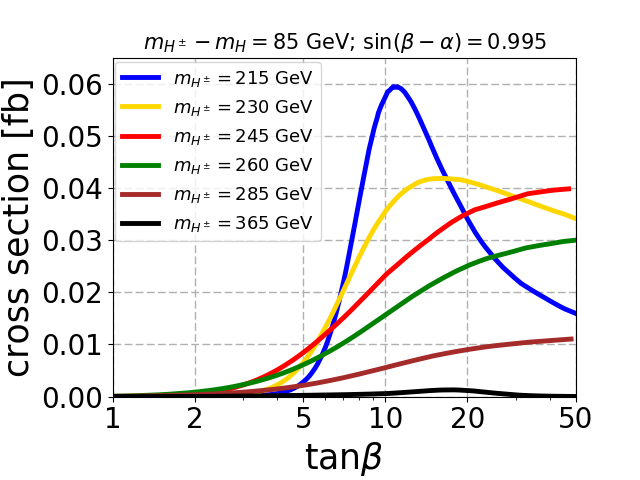}
\end{center}
\begin{center}
\includegraphics[width=77mm]{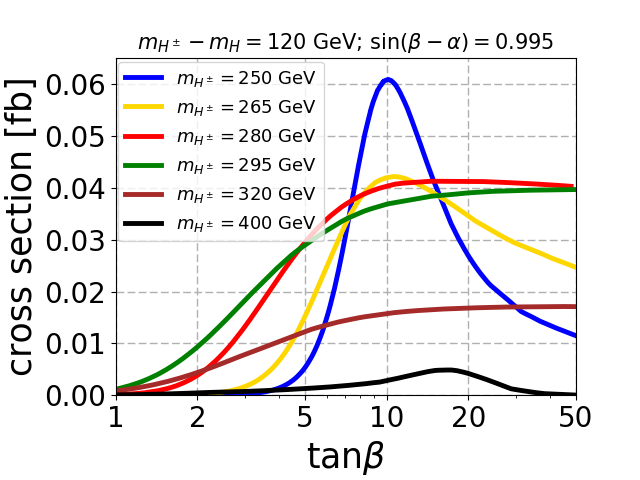}
\end{center}
\caption{The same sign trilepton (SS3L) signal cross-sections after all cuts is plotted against $\tan\beta$ for different charged Higgs masses with three mass differences, 60 GeV (top left) 85 GeV (top right), 120 GeV (bottom) and $\sin(\beta - \alpha) = 0.995$. Please refer to text for the explanation of different curves.
}
\label{fig:signal-x-sections}
\end{figure}

To set the exclusion limits on our signal, we compute the significance ($\sigma_{exc}$) for exclusion using a  likelihood ratio 
method~\cite{Cowan:2010js} given by: 
\begin{equation}
\sigma_{exc} = \sqrt{-2\ln\Big(\frac{L(S+B|B)}{L(B|B)}\Big)}~, 
\hspace{1cm}\text{where}~~~L(x|n) = \frac{x^n}{n!}e^{-x}.
\end{equation}
To estimate the discovery reach, we compute the following significance,
\begin{eqnarray}
\sigma_{dis} &=& \sqrt{-2 \ln\Big(\frac{L(B|S + B)}{L(S + B| S + B)}\Big)}. \nonumber
\end{eqnarray}
For exclusion we demand $\sigma_{exc} \geq 2$ and for discovery we demand $\sigma_{dis} \geq 5$. 
In Fig~\ref{fig:signal-x-section-lumi} we shows the signal cross-sections required for $2\sigma$ exclusion 
and $5\sigma$ discovery as a function of integrated luminosity. 

After applying the selection cuts, the same sign trilepton signal cross-section for  different charged Higgs 
mass is shown in Fig.~\ref{fig:signal-x-sections} for a mass gap of 60 GeV(top left), 85 GeV(top right) and 120 GeV (bottom). 
For all the plots we fix $\sin(\beta-\alpha) = 0.995$, and the fermiophobic limit for such $\sin(\beta-\alpha)$ the 
fermiophobic limit happens at $\tb=10$ as described in Sec.~\ref{subsec:model} and manifest in the top panel of 
Fig.~\ref{fig:Branching ratios} for $m_H=130$ GeV. This is prominent when mass of $\hpm$  and in turn mass of $H$ 
is the lowest and depicted in blue curve. 

As the mass of the BSM  heavy Higgs $H$ increases, the $BR(H\to W^+W^-)$ remains large for any $\tb$ and the cross-section 
remains large beyond the fermiophobic region, which is depicted in red and green curves in Fig.~\ref{fig:signal-x-sections}. 
Notice that for the mass gap of 60 GeV, the cross-section shown by yellow curve for $m_\hpm = 205$ GeV is large at high 
$\tb$ compared to the blue curve for $m_\hpm = 190$ GeV. This is because the $BR(\hpm \to H W^\pm)$ is sizable only for $\tb \gg 10$.
When the mass of $\hpm$ and therefore $H$ is least, $BR(H\to W^+W^-)$ peaks at $\tb=10$ where $BR(\hpm \to H W^\pm)$ is low resulting to a low signal cross-section. This is the case for $m_\hpm = 190$ GeV as shown in blue curve. When mass of $\hpm$ increases, $BR(\hpm \to H W^\pm)$
saturates beyond $\tb = 10$, and consequently $BR(\hpm \to H W^\pm)$ becomes large at high $\tan\beta$ which pushes the yellow curve beyond the blue curve.
For the mass gap of 120 GeV, the $BR(\hpm(A) \to H W^\pm(Z))$ saturates at a lower $\tb$ compared to the mass gaps of 60 GeV and 85 GeV. Hence the signal cross-section also saturates at a lower $\tan\beta$. This is clear from the red, green and brown curves of the three mass gaps. The same argument explains why the width of the blue curve, which corresponds to $m_H = 130$ GeV is minimum for the mass gap of 120 GeV and maximum for the mass gap of 60 GeV.  Finally, when the $H\to hh$ decay comes into play, the signal cross-section decreases rapidly as shown by the black curves for all the mass gaps.

\begin{figure}[t!]
\begin{center}
\includegraphics[width=75mm]{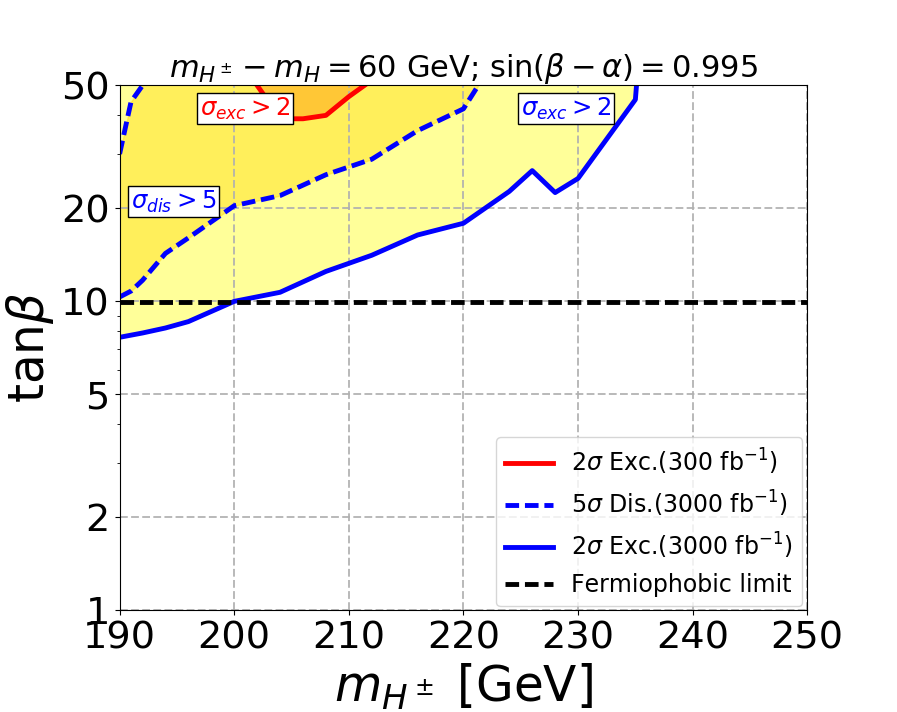}
\includegraphics[width=79mm]{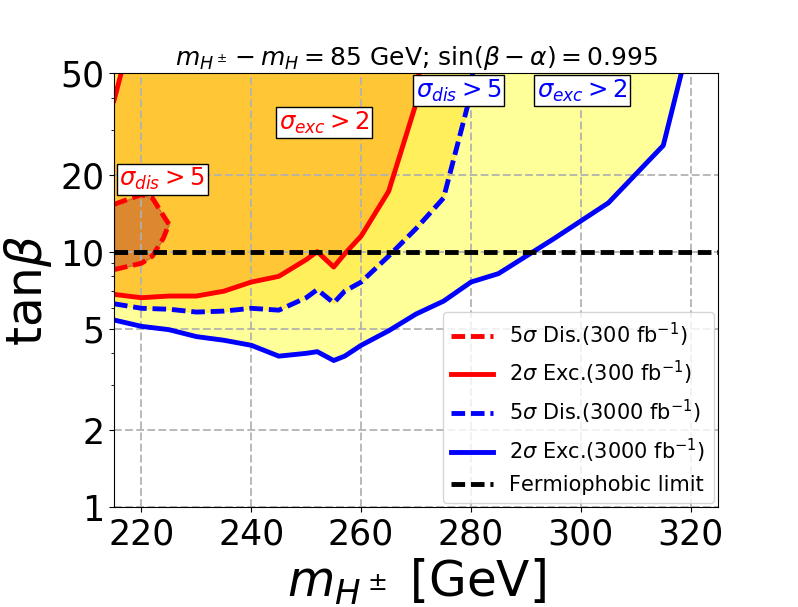}
\end{center}
\begin{center}
\includegraphics[width=77mm]{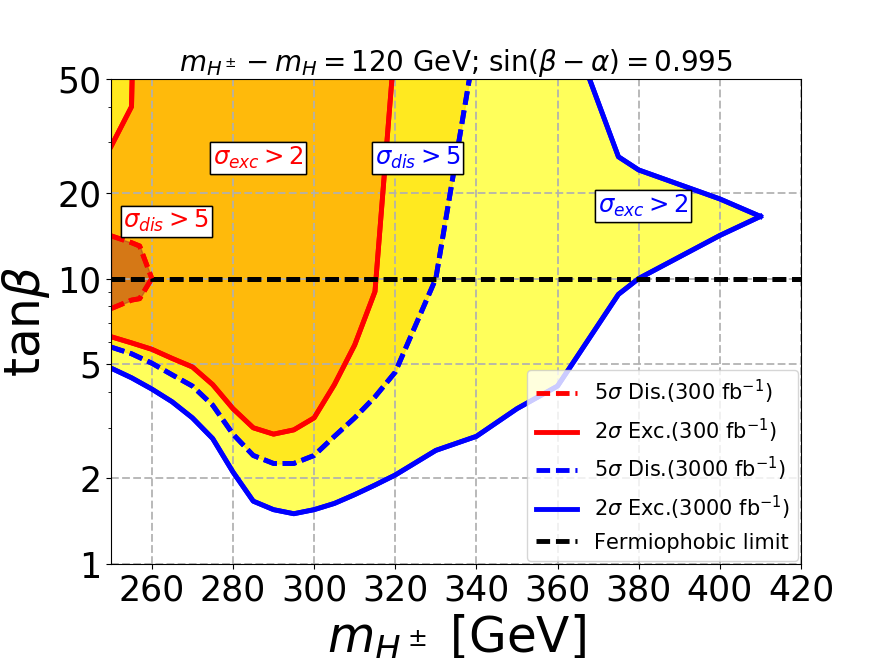}
\end{center}
\caption{The exclusion and discovery limits on the $(m_{\hpm} - \tan\beta)$ parameter space for 300 $fb{^{-1}}$ 
and 3000 $fb{^{-1}}$ lumiosities at $\sqrt{s} = 13$ TeV for three mass differences, 60 GeV (top left) 85 GeV (top right) and 120 GeV (bottom) 
are given. The parameter $\sin(\beta - \alpha)=0.995$ leads to the fermiophobic limit of the heavy Higgs at 
$\tan\beta \sim 10$ as shown by the black dashed lines.} 
\label{fig:exclusion995}
\end{figure}

Now we will show our main result of exclusion and discovery region in $m_{H^\pm}-\tb$ plane for the different 
mass gaps. In Fig.~\ref{fig:exclusion995} we displayed the reach of LHC for three mass gaps where $\sin(\beta-\alpha) = 0.995$. 
The fermiophobic limit corresponds to $\tb=10$ and 
is shown in the horizontal black dashed line. The exclusion(discovery) contours are depicted by solid(dashed) 
curves and red(blue) curves corresponds to integrated luminosity of 300(3000)$fb^{-1}$. 

For the mass gap of 60 GeV, the low $BR(\hpm \to H W^\pm)$ at the fermiophobic limit of $\tb=10$ makes the signal cross-section low as mentioned before. Hence the signal is 
not enough for discovery at 300 $fb^{-1}$ as seen in top left figure.
For the mass gaps of 85 GeV and 120 GeV, the discovery regions at 300$fb^{-1}$ are confined only for low $H^\pm$ mass where the signal cross-sections are large. We also observe a dip in the discovery regions at 3000$fb^{-1}$ and exclusion regions at 300$fb^{-1}$ and 3000$fb^{-1}$ due to the enhancement to the signal coming from the on-shell $H\to W^+W^-$ mode. The on-shell effect can also be seen for mass gap of 60 GeV by the presence of a dip in the exclusion region at 3000$fb^{-1}$.   
From the plots, it is 
evident that the SS3L signal is capable of excluding a substantial parameter space at the large $\tb$ region. 
As the mass gap increases, the $\hpm$ decay to $W^\pm H$ saturates at a 
lower value of $\tb$, which pushes the limits to lower $\tb$ region and thus excludes even larger parameter space. This is most evident for the mass gap of 120 GeV as shown in the bottom panel. Also for the mass gap of 120 GeV, when the mass of $\hpm$ is large, we can not exclude the large $\tb$ region as the decay $H\to hh$ 
dominates at large $\tb$, and our signal becomes irrelevant.

\begin{figure}[t!]
\begin{center}
\includegraphics[width=77mm]{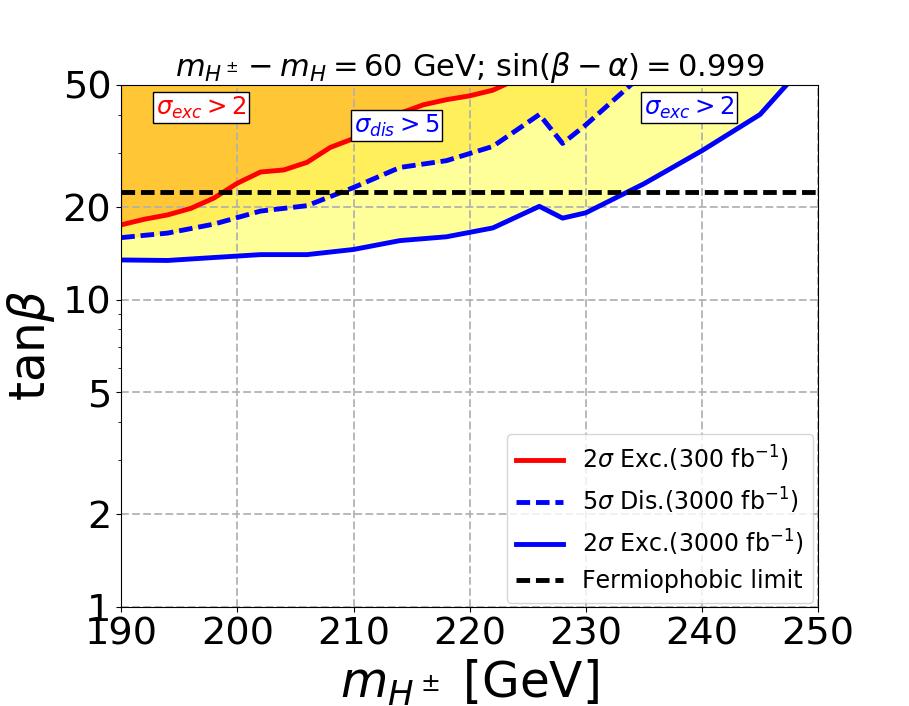}
\includegraphics[width=77mm]{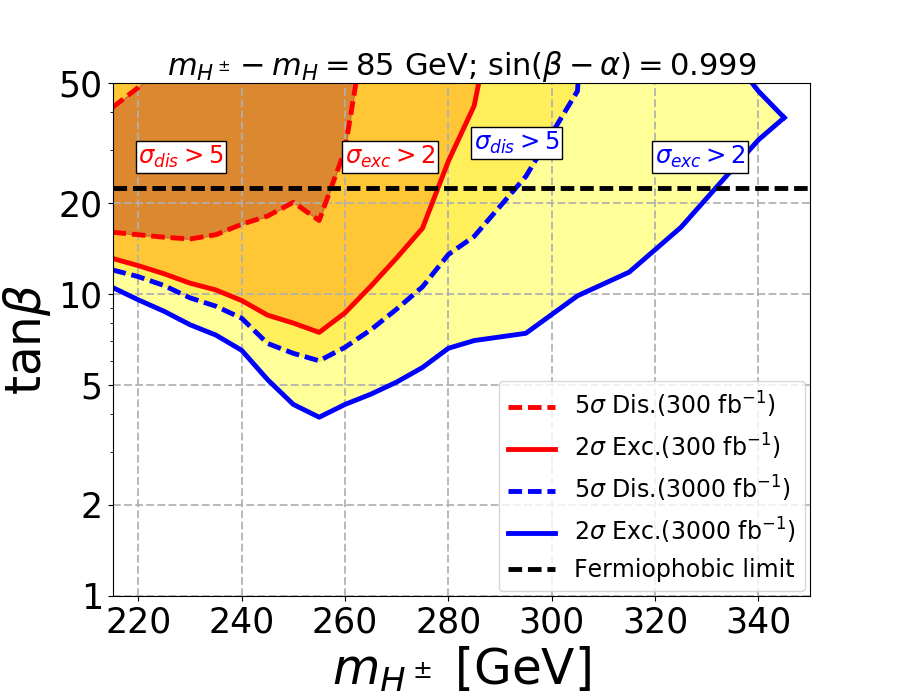}
\end{center}
\begin{center}
\includegraphics[width=77mm]{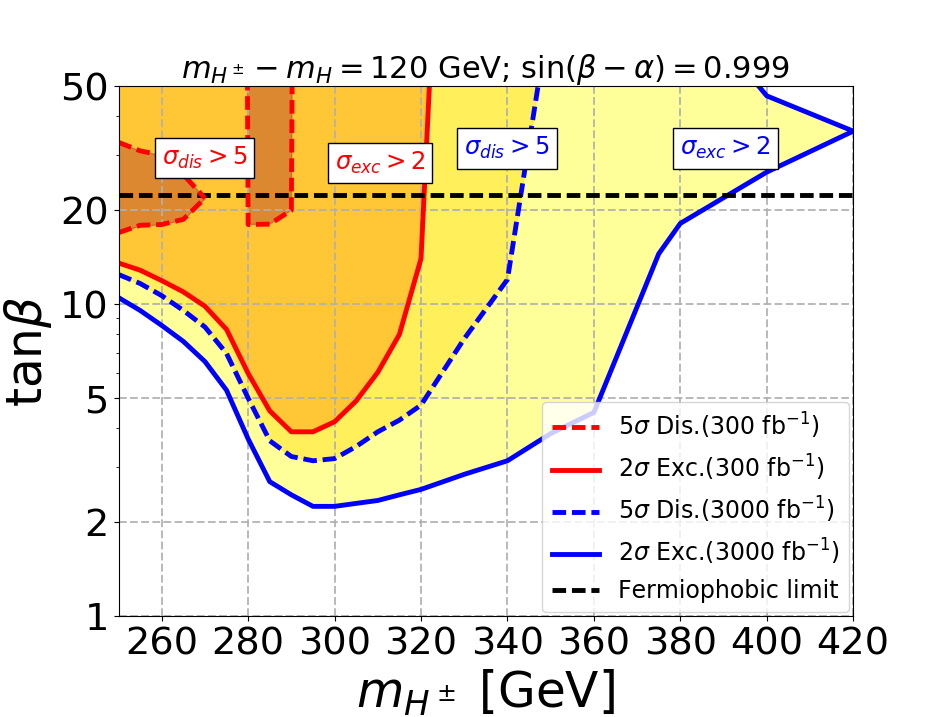}
\end{center}
\caption{Same as Fig.~\ref{fig:exclusion995} with $\sin(\beta - \alpha)=0.999$.} 
\label{fig:exclusion999}
\end{figure}


To show how the proposed signal works even closer to the alignment limit, we have performed the phenomenological 
analysis for $\sin(\beta-\alpha) =0.999$. The limit becomes stronger in this scenario because of the following reasons:
\begin{itemize}
 \item The $H^\pm H W^\mp $ coupling is proportional to $\sin(\beta-\alpha)$, the production cross-section and the 
 branching ratio of $\hpm$ increases in this case. 
 \item $H\to W^+W^-$ remains dominant even with $\cos(\beta-\alpha)$ suppression since the fermionic decay modes of $H$ 
is $\tb$ suppressed.
\item The subdominant signal is enhanced as the $HAZ$ coupling is proportional to $\sin(\beta - \alpha)$.
\item $\lambda_{Hhh}$, being proportional to $\cos(\beta-\alpha)$ becomes small and 
$H\to hh$ decay is suppressed when allowed. 
\end{itemize}
 Effectively, the exclusion and discovery limits become stronger for $\sin(\beta-\alpha) =0.999$. 
This is evident from Fig.~\ref{fig:exclusion999}, specially for the mass gap of 60 GeV the exclusion region at 300 $fb^{-1}$ is much more enhanced. Also the discovery regions at 300 $fb^{-1}$ for the mass gap of 85 GeV and 120 GeV are more enhanced. The discovery region at 300 $fb^{-1}$ for the mass gap of 120 GeV shows a discontinuity as for small $m_{H^\pm}$ the cross-section is large and it again reappears when $H$ decay to on-shell $W$ pair is open. Just like the case of $\sin(\beta -\alpha)=0.995$, the limits become most stringent when the on-shell decay of $H\to W^+W^-$ opens up showing a dip around 
$m_H=165$ GeV. In Fig.~\ref{fig:exclusion999} the fermiophobic limit corresponds to $\tb\simeq22$. Since the value of $\tb$ for fermiophobic limit has moved upwards, we expect the 
exclusion and discovery regions to shift to higher $\tan\beta$, which is quite visible for the discovery 
regions at 300 $fb^{-1}$ for the mass gaps of 85 GeV and 120 GeV. Similarly the discovery and exclusion regions at 3000 $fb^{-1}$ for the mass gap of 60 GeV are shifted upwards. However, the overall signal cross-sections have increased due to the 
enhancement of $\sin(\beta-\alpha)$, the shift of the discovery regions at 3000 $fb^{-1}$ and exclusion regions at 300 $fb^{-1}$ and 3000 $fb^{-1}$ for the mass gap of 85 GeV and 120 GeV are negligible.

\section{Conclusion}\label{sec:conclusion}

Observation of a charged Higgs at the LHC will indicate the presence of an extended Higgs sector. 
The search strategies of looking for a charged Higgs at the LHC dominantly depends on the $\hpm tb$ 
coupling. However, the $\hpm tb$ coupling can be small for a fermiophobic $\hpm$ scenario like the 
type-I 2HDM model. As a result, the limit on a charged Higgs is non-existent unless $\tb$ is close 
to unity.  Here we proposed the same sign trilepton signal to complement the existing searches. 
The SS3L signature appears when a charged Higgs is produced via electroweak interaction in association 
with $H$ or $A$ and subsequently decays to heavy gauge bosons. By performing a detailed phenomenological 
analysis, we demonstrate that our proposed signal is capable of extending the reach of LHC up to a 
very high $\tb$ region for charged Higgs mass of up to 400 GeV. 
The decay of the BSM scalars $\hpm, A$ and $H$ depends on the mass gap among themselves and to cover a large 
model parameter space we studied three mass gaps between $H^\pm$ and $H$, viz. 60 GeV, 85 GeV and 120 GeV, which covers complete
off-shell to fully on-shell decay of various BSM scalars.
We also showed that the dependency 
of the signal on mixing angle $\sin(\beta-\alpha)$ is relatively weak, and a slight deviation from 
the alignment limit is enough to explore the SS3L signature. If the deviation from the alignment 
limit is significant, then the proposed signal will not work efficiently as both the couplings 
$H^\pm H W^\mp $ and $HAZ$ couplings are proportional to $\sin(\beta-\alpha)$. Also, the exact 
fermiophobic limit will move towards the low $\tb$ region, which is already constrained via the 
$\hpm$ search in association with the top quark.

\section*{Acknowledgments}
The authors would like to thank Ravindra Kumar Verma for some useful comments and discussions. T.M. was supported by a KIAS Individual Grant PG073502 at Korea Institute for Advanced Study. P.S. was supported by the appointment to the JRG Program at the APCTP through the Science and Technology Promotion Fund and Lottery Fund of the Korean Government. This was also supported by the Korean Local Governments - Gyeongsangbuk-do Province and Pohang City.

 \bibliographystyle{JHEP}
 \bibliography{2hdm}

\providecommand{\href}[2]{#2}\begingroup\raggedright\begin{thebibliography}{10}

\bibitem{Gunion:1989we}
J.~F. Gunion, H.~E. Haber, G.~L. Kane and S.~Dawson, \emph{{The Higgs Hunter's
  Guide}}, vol.~80. 2000.

\bibitem{Djouadi:2005gj}
A.~Djouadi, \emph{{The Anatomy of electro-weak symmetry breaking. II. The Higgs
  bosons in the minimal supersymmetric model}},
  \href{https://doi.org/10.1016/j.physrep.2007.10.005}{\emph{Phys. Rept.}
  {\bfseries 459} (2008) 1}
  [\href{https://arxiv.org/abs/hep-ph/0503173}{{\ttfamily hep-ph/0503173}}].

\bibitem{Branco:2011iw}
G.~C. Branco, P.~M. Ferreira, L.~Lavoura, M.~N. Rebelo, M.~Sher and J.~P.
  Silva, \emph{{Theory and phenomenology of two-Higgs-doublet models}},
  \href{https://doi.org/10.1016/j.physrep.2012.02.002}{\emph{Phys. Rept.}
  {\bfseries 516} (2012) 1} [\href{https://arxiv.org/abs/1106.0034}{{\ttfamily
  1106.0034}}].

\bibitem{ATLAS:2013uxj}
{\scshape ATLAS} collaboration, \emph{{Search for a light charged Higgs boson
  in the decay channel $H^+ \to c\bar{s}$ in $t\bar{t}$ events using pp
  collisions at $\sqrt{s}$ = 7 TeV with the ATLAS detector}},
  \href{https://doi.org/10.1140/epjc/s10052-013-2465-z}{\emph{Eur. Phys. J. C}
  {\bfseries 73} (2013) 2465}
  [\href{https://arxiv.org/abs/1302.3694}{{\ttfamily 1302.3694}}].

\bibitem{ATLAS:2014otc}
{\scshape ATLAS} collaboration, \emph{{Search for charged Higgs bosons decaying
  via $H^{\pm} \rightarrow \tau^{\pm}\nu$ in fully hadronic final states using
  $pp$ collision data at $\sqrt{s} = 8$ TeV with the ATLAS detector}},
  \href{https://doi.org/10.1007/JHEP03(2015)088}{\emph{JHEP} {\bfseries 03}
  (2015) 088} [\href{https://arxiv.org/abs/1412.6663}{{\ttfamily 1412.6663}}].

\bibitem{CMS:2015lsf}
{\scshape CMS} collaboration, \emph{{Search for a charged Higgs boson in pp
  collisions at $ \sqrt{s}=8 $ TeV}},
  \href{https://doi.org/10.1007/JHEP11(2015)018}{\emph{JHEP} {\bfseries 11}
  (2015) 018} [\href{https://arxiv.org/abs/1508.07774}{{\ttfamily
  1508.07774}}].

\bibitem{CMS:2015yvc}
{\scshape CMS} collaboration, \emph{{Search for a light charged Higgs boson
  decaying to $ \mathrm{c}\overline{\mathrm{s}} $ in pp collisions at $
  \sqrt{s}=8 $ TeV}},
  \href{https://doi.org/10.1007/JHEP12(2015)178}{\emph{JHEP} {\bfseries 12}
  (2015) 178} [\href{https://arxiv.org/abs/1510.04252}{{\ttfamily
  1510.04252}}].

\bibitem{ATLAS:2016avi}
{\scshape ATLAS} collaboration, \emph{{Search for charged Higgs bosons produced
  in association with a top quark and decaying via $H^{\pm} \rightarrow
  \tau\nu$ using $pp$ collision data recorded at $\sqrt{s} = 13$ TeV by the
  ATLAS detector}},
  \href{https://doi.org/10.1016/j.physletb.2016.06.017}{\emph{Phys. Lett. B}
  {\bfseries 759} (2016) 555}
  [\href{https://arxiv.org/abs/1603.09203}{{\ttfamily 1603.09203}}].

\bibitem{ATLAS:2018gfm}
{\scshape ATLAS} collaboration, \emph{{Search for charged Higgs bosons decaying
  via $H^{\pm} \to \tau^{\pm}\nu_{\tau}$ in the $\tau$+jets and $\tau$+lepton
  final states with 36 fb$^{-1}$ of $pp$ collision data recorded at $\sqrt{s} =
  13$ TeV with the ATLAS experiment}},
  \href{https://doi.org/10.1007/JHEP09(2018)139}{\emph{JHEP} {\bfseries 09}
  (2018) 139} [\href{https://arxiv.org/abs/1807.07915}{{\ttfamily
  1807.07915}}].

\bibitem{ATLAS:2018ntn}
{\scshape ATLAS} collaboration, \emph{{Search for charged Higgs bosons decaying
  into top and bottom quarks at $\sqrt{s}$ = 13 TeV with the ATLAS detector}},
  \href{https://doi.org/10.1007/JHEP11(2018)085}{\emph{JHEP} {\bfseries 11}
  (2018) 085} [\href{https://arxiv.org/abs/1808.03599}{{\ttfamily
  1808.03599}}].

\bibitem{CMS:2018dzl}
{\scshape CMS} collaboration, \emph{{Search for a charged Higgs boson decaying
  to charm and bottom quarks in proton-proton collisions at $ \sqrt{s}=8 $
  TeV}}, \href{https://doi.org/10.1007/JHEP11(2018)115}{\emph{JHEP} {\bfseries
  11} (2018) 115} [\href{https://arxiv.org/abs/1808.06575}{{\ttfamily
  1808.06575}}].

\bibitem{CMS:2019bfg}
{\scshape CMS} collaboration, \emph{{Search for charged Higgs bosons in the
  H$^{\pm}$ $\to$ $\tau^{\pm}\nu_\tau$ decay channel in proton-proton
  collisions at $\sqrt{s} =$ 13 TeV}},
  \href{https://doi.org/10.1007/JHEP07(2019)142}{\emph{JHEP} {\bfseries 07}
  (2019) 142} [\href{https://arxiv.org/abs/1903.04560}{{\ttfamily
  1903.04560}}].

\bibitem{ATLAS:2020jqj}
{\scshape ATLAS} collaboration, \emph{{Search for charged Higgs bosons decaying
  into a top-quark and a bottom-quark at $\sqrt s$ = 13 TeV with the ATLAS
  detector}}, .

\bibitem{ATLAS:2021upq}
{\scshape ATLAS} collaboration, \emph{{Search for charged Higgs bosons decaying
  into a top quark and a bottom quark at $ \sqrt{\mathrm{s}} $ = 13 TeV with
  the ATLAS detector}},
  \href{https://doi.org/10.1007/JHEP06(2021)145}{\emph{JHEP} {\bfseries 06}
  (2021) 145} [\href{https://arxiv.org/abs/2102.10076}{{\ttfamily
  2102.10076}}].

\bibitem{CMS:2017fgp}
{\scshape CMS} collaboration, \emph{{Search for Charged Higgs Bosons Produced
  via Vector Boson Fusion and Decaying into a Pair of $W$ and $Z$ Bosons Using
  $pp$ Collisions at $\sqrt{s}=13\text{ }\text{ }\mathrm{TeV}$}},
  \href{https://doi.org/10.1103/PhysRevLett.119.141802}{\emph{Phys. Rev. Lett.}
  {\bfseries 119} (2017) 141802}
  [\href{https://arxiv.org/abs/1705.02942}{{\ttfamily 1705.02942}}].

\bibitem{ATLAS:2018iui}
{\scshape ATLAS} collaboration, \emph{{Search for resonant $WZ$ production in
  the fully leptonic final state in proton-proton collisions at $\sqrt{s} = 13$
  TeV with the ATLAS detector}},
  \href{https://doi.org/10.1016/j.physletb.2018.10.021}{\emph{Phys. Lett. B}
  {\bfseries 787} (2018) 68}
  [\href{https://arxiv.org/abs/1806.01532}{{\ttfamily 1806.01532}}].

\bibitem{Coleppa_2020}
B.~Coleppa, A.~Sarkar and S.~K. Rai, \emph{Charged higgs boson discovery
  prospects}, \href{https://doi.org/10.1103/physrevd.101.055030}{\emph{Physical
  Review D} {\bfseries 101} (2020) }.

\bibitem{Coleppa:2021wjx}
B.~Coleppa, G.~B. Krishna and A.~Sarkar, \emph{{Charged Higgs Prospects In
  Extended Gauge Models}},  \href{https://arxiv.org/abs/2107.03993}{{\ttfamily
  2107.03993}}.

\bibitem{Chen:2019pkq}
N.~Chen, T.~Han, S.~Li, S.~Su, W.~Su and Y.~Wu, \emph{{Type-I 2HDM under the
  Higgs and Electroweak Precision Measurements}},
  \href{https://doi.org/10.1007/JHEP08(2020)131}{\emph{JHEP} {\bfseries 08}
  (2020) 131} [\href{https://arxiv.org/abs/1912.01431}{{\ttfamily
  1912.01431}}].

\bibitem{Kling:2020hmi}
F.~Kling, S.~Su and W.~Su, \emph{{2HDM Neutral Scalars under the LHC}},
  \href{https://doi.org/10.1007/JHEP06(2020)163}{\emph{JHEP} {\bfseries 06}
  (2020) 163} [\href{https://arxiv.org/abs/2004.04172}{{\ttfamily
  2004.04172}}].

\bibitem{Coleppa:2014cca}
B.~Coleppa, F.~Kling and S.~Su, \emph{{Charged Higgs search via $AW^\pm/HW^\pm$
  channel}}, \href{https://doi.org/10.1007/JHEP12(2014)148}{\emph{JHEP}
  {\bfseries 12} (2014) 148} [\href{https://arxiv.org/abs/1408.4119}{{\ttfamily
  1408.4119}}].

\bibitem{Enberg:2014pua}
R.~Enberg, W.~Klemm, S.~Moretti, S.~Munir and G.~Wouda, \emph{{Charged Higgs
  boson in the $W^\pm$ Higgs channel at the Large Hadron Collider}},
  \href{https://doi.org/10.1016/j.nuclphysb.2015.02.001}{\emph{Nucl. Phys. B}
  {\bfseries 893} (2015) 420}
  [\href{https://arxiv.org/abs/1412.5814}{{\ttfamily 1412.5814}}].

\bibitem{Kling:2015uba}
F.~Kling, A.~Pyarelal and S.~Su, \emph{{Light Charged Higgs Bosons to AW/HW via
  Top Decay}}, \href{https://doi.org/10.1007/JHEP11(2015)051}{\emph{JHEP}
  {\bfseries 11} (2015) 051}
  [\href{https://arxiv.org/abs/1504.06624}{{\ttfamily 1504.06624}}].

\bibitem{Akeroyd:2016ymd}
A.~G. Akeroyd et~al., \emph{{Prospects for charged Higgs searches at the LHC}},
  \href{https://doi.org/10.1140/epjc/s10052-017-4829-2}{\emph{Eur. Phys. J. C}
  {\bfseries 77} (2017) 276}
  [\href{https://arxiv.org/abs/1607.01320}{{\ttfamily 1607.01320}}].

\bibitem{Arhrib:2016wpw}
A.~Arhrib, R.~Benbrik and S.~Moretti, \emph{{Bosonic Decays of Charged Higgs
  Bosons in a 2HDM Type-I}},
  \href{https://doi.org/10.1140/epjc/s10052-017-5197-7}{\emph{Eur. Phys. J. C}
  {\bfseries 77} (2017) 621}
  [\href{https://arxiv.org/abs/1607.02402}{{\ttfamily 1607.02402}}].

\bibitem{Alves:2017snd}
D.~S.~M. Alves, S.~El~Hedri, A.~M. Taki and N.~Weiner, \emph{{Charged Higgs
  Signals in $t\,\overline{t}\,H$ Searches}},
  \href{https://doi.org/10.1103/PhysRevD.96.075032}{\emph{Phys. Rev. D}
  {\bfseries 96} (2017) 075032}
  [\href{https://arxiv.org/abs/1703.06834}{{\ttfamily 1703.06834}}].

\bibitem{Arhrib:2019ywg}
A.~Arhrib, K.~Cheung and C.-T. Lu, \emph{{Same-sign charged Higgs boson pair
  production in bosonic decay channels at the HL-LHC and HE-LHC}},
  \href{https://doi.org/10.1103/PhysRevD.102.095026}{\emph{Phys. Rev. D}
  {\bfseries 102} (2020) 095026}
  [\href{https://arxiv.org/abs/1910.02571}{{\ttfamily 1910.02571}}].

\bibitem{Arhrib:2020tqk}
A.~Arhrib, R.~Benbrik, H.~Harouiz, S.~Moretti, Y.~Wang and Q.-S. Yan,
  \emph{{Implications of a light charged Higgs boson at the LHC run III in the
  2HDM}}, \href{https://doi.org/10.1103/PhysRevD.102.115040}{\emph{Phys. Rev.
  D} {\bfseries 102} (2020) 115040}
  [\href{https://arxiv.org/abs/2003.11108}{{\ttfamily 2003.11108}}].

\bibitem{Sanyal:2019xcp}
P.~Sanyal, \emph{{Limits on the Charged Higgs Parameters in the Two Higgs
  Doublet Model using CMS $\sqrt{s}=13$ TeV Results}},
  \href{https://doi.org/10.1140/epjc/s10052-019-7431-y}{\emph{Eur. Phys. J. C}
  {\bfseries 79} (2019) 913}
  [\href{https://arxiv.org/abs/1906.02520}{{\ttfamily 1906.02520}}].

\bibitem{Akeroyd_1999}
A.~Akeroyd, \emph{Three-body decays of higgs bosons at lep2 and application to
  a hidden fermiophobic higgs},
  \href{https://doi.org/10.1016/s0550-3213(98)00845-1}{\emph{Nuclear Physics B}
  {\bfseries 544} (1999) 557–575}.

\bibitem{Demirci_2020}
M.~Demirci, \emph{Precise predictions for charged higgs boson pair production
  in photon-photon collisions},
  \href{https://doi.org/10.1016/j.nuclphysb.2020.115235}{\emph{Nuclear Physics
  B} {\bfseries 961} (2020) 115235}.

\bibitem{Kanemura:2001hz}
S.~Kanemura and C.~P. Yuan, \emph{{Testing supersymmetry in the associated
  production of CP odd and charged Higgs bosons}},
  \href{https://doi.org/10.1016/S0370-2693(02)01321-7}{\emph{Phys. Lett. B}
  {\bfseries 530} (2002) 188}
  [\href{https://arxiv.org/abs/hep-ph/0112165}{{\ttfamily hep-ph/0112165}}].

\bibitem{Cao:2003tr}
Q.-H. Cao, S.~Kanemura and C.~P. Yuan, \emph{{Associated production of CP odd
  and charged Higgs bosons at hadron colliders}},
  \href{https://doi.org/10.1103/PhysRevD.69.075008}{\emph{Phys. Rev. D}
  {\bfseries 69} (2004) 075008}
  [\href{https://arxiv.org/abs/hep-ph/0311083}{{\ttfamily hep-ph/0311083}}].

\bibitem{Belyaev:2006rf}
A.~Belyaev, Q.-H. Cao, D.~Nomura, K.~Tobe and C.~P. Yuan, \emph{{Light MSSM
  Higgs boson scenario and its test at hadron colliders}},
  \href{https://doi.org/10.1103/PhysRevLett.100.061801}{\emph{Phys. Rev. Lett.}
  {\bfseries 100} (2008) 061801}
  [\href{https://arxiv.org/abs/hep-ph/0609079}{{\ttfamily hep-ph/0609079}}].

\bibitem{Chun:2018vsn}
E.~J. Chun, S.~Dwivedi, T.~Mondal, B.~Mukhopadhyaya and S.~K. Rai,
  \emph{{Reconstructing heavy Higgs boson masses in a type X two-Higgs-doublet
  model with a light pseudoscalar particle}},
  \href{https://doi.org/10.1103/PhysRevD.98.075008}{\emph{Phys. Rev.}
  {\bfseries D98} (2018) 075008}
  [\href{https://arxiv.org/abs/1807.05379}{{\ttfamily 1807.05379}}].

\bibitem{Bahl:2021str}
H.~Bahl, T.~Stefaniak and J.~Wittbrodt, \emph{{The forgotten channels: charged
  Higgs boson decays to a W$^{±}$ and a non-SM-like Higgs boson}},
  \href{https://doi.org/10.1007/JHEP06(2021)183}{\emph{JHEP} {\bfseries 06}
  (2021) 183} [\href{https://arxiv.org/abs/2103.07484}{{\ttfamily
  2103.07484}}].

\bibitem{Arhrib:2017wmo}
A.~Arhrib, R.~Benbrik, R.~Enberg, W.~Klemm, S.~Moretti and S.~Munir,
  \emph{{Identifying a light charged Higgs boson at the LHC Run II}},
  \href{https://doi.org/10.1016/j.physletb.2017.10.006}{\emph{Phys. Lett. B}
  {\bfseries 774} (2017) 591}
  [\href{https://arxiv.org/abs/1706.01964}{{\ttfamily 1706.01964}}].

\bibitem{Enberg:2018pye}
R.~Enberg, W.~Klemm, S.~Moretti and S.~Munir, \emph{{Electroweak production of
  multiple (pseudo)scalars in the 2HDM}},
  \href{https://doi.org/10.1140/epjc/s10052-019-7025-8}{\emph{Eur. Phys. J. C}
  {\bfseries 79} (2019) 512}
  [\href{https://arxiv.org/abs/1812.01147}{{\ttfamily 1812.01147}}].

\bibitem{Enberg:2018nfv}
R.~Enberg, W.~Klemm, S.~Moretti and S.~Munir, \emph{{Signatures of the Type-I
  2HDM at the LHC}}, \href{https://doi.org/10.22323/1.347.0013}{\emph{PoS}
  {\bfseries CORFU2018} (2018) 013}
  [\href{https://arxiv.org/abs/1812.08623}{{\ttfamily 1812.08623}}].

\bibitem{Arhrib:2021xmc}
A.~Arhrib, R.~Benbrik, M.~Krab, B.~Manaut, S.~Moretti, Y.~Wang et~al.,
  \emph{{New Discovery Modes for a Light Charged Higgs Boson at the LHC}},
  \href{https://arxiv.org/abs/2106.13656}{{\ttfamily 2106.13656}}.

\bibitem{Wang:2021pxc}
Y.~Wang, A.~Arhrib, R.~Benbrik, M.~Krab, B.~Manaut, S.~Moretti et~al.,
  \emph{{Analysis of $W^\pm+4\gamma$ in the 2HDM Type-I at the LHC}},
  \href{https://arxiv.org/abs/2107.01451}{{\ttfamily 2107.01451}}.

\bibitem{Akeroyd:2003bt}
A.~G. Akeroyd and M.~A. Diaz, \emph{{Searching for a light fermiophobic Higgs
  boson at the Tevatron}},
  \href{https://doi.org/10.1103/PhysRevD.67.095007}{\emph{Phys. Rev. D}
  {\bfseries 67} (2003) 095007}
  [\href{https://arxiv.org/abs/hep-ph/0301203}{{\ttfamily hep-ph/0301203}}].

\bibitem{Akeroyd_2004}
A.~G. Akeroyd, M.~A. Díaz and F.~J. Pacheco, \emph{Double fermiophobic higgs
  boson production at the cern lhc and a linear collider},
  \href{https://doi.org/10.1103/physrevd.70.075002}{\emph{Physical Review D}
  {\bfseries 70} (2004) }.

\bibitem{PhysRevD.77.115013}
A.~Arhrib, R.~Benbrik and C.-W. Chiang, \emph{Probing triple higgs couplings of
  the two higgs doublet model at a linear collider},
  \href{https://doi.org/10.1103/PhysRevD.77.115013}{\emph{Phys. Rev. D}
  {\bfseries 77} (2008) 115013}.

\bibitem{Gunion:2002zf}
J.~F. Gunion and H.~E. Haber, \emph{{The CP conserving two Higgs doublet model:
  The Approach to the decoupling limit}},
  \href{https://doi.org/10.1103/PhysRevD.67.075019}{\emph{Phys. Rev.}
  {\bfseries D67} (2003) 075019}
  [\href{https://arxiv.org/abs/hep-ph/0207010}{{\ttfamily hep-ph/0207010}}].

\bibitem{Kanemura:1993hm}
S.~Kanemura, T.~Kubota and E.~Takasugi, \emph{{Lee-Quigg-Thacker bounds for
  Higgs boson masses in a two doublet model}},
  \href{https://doi.org/10.1016/0370-2693(93)91205-2}{\emph{Phys. Lett. B}
  {\bfseries 313} (1993) 155}
  [\href{https://arxiv.org/abs/hep-ph/9303263}{{\ttfamily hep-ph/9303263}}].

\bibitem{Akeroyd:2000wc}
A.~G. Akeroyd, A.~Arhrib and E.-M. Naimi, \emph{{Note on tree level unitarity
  in the general two Higgs doublet model}},
  \href{https://doi.org/10.1016/S0370-2693(00)00962-X}{\emph{Phys. Lett. B}
  {\bfseries 490} (2000) 119}
  [\href{https://arxiv.org/abs/hep-ph/0006035}{{\ttfamily hep-ph/0006035}}].

\bibitem{Eriksson:2009ws}
D.~Eriksson, J.~Rathsman and O.~Stal, \emph{{2HDMC: Two-Higgs-Doublet Model
  Calculator Physics and Manual}},
  \href{https://doi.org/10.1016/j.cpc.2009.09.011}{\emph{Comput. Phys. Commun.}
  {\bfseries 181} (2010) 189}
  [\href{https://arxiv.org/abs/0902.0851}{{\ttfamily 0902.0851}}].

\bibitem{Peskin:1991sw}
M.~E. Peskin and T.~Takeuchi, \emph{{Estimation of oblique electroweak
  corrections}}, \href{https://doi.org/10.1103/PhysRevD.46.381}{\emph{Phys.
  Rev. D} {\bfseries 46} (1992) 381}.

\bibitem{Bechtle:2013wla}
P.~Bechtle, O.~Brein, S.~Heinemeyer, O.~St\r{a}l, T.~Stefaniak, G.~Weiglein
  et~al., \emph{{$\mathsf{HiggsBounds}-4$: Improved Tests of Extended Higgs
  Sectors against Exclusion Bounds from LEP, the Tevatron and the LHC}},
  \href{https://doi.org/10.1140/epjc/s10052-013-2693-2}{\emph{Eur. Phys. J. C}
  {\bfseries 74} (2014) 2693}
  [\href{https://arxiv.org/abs/1311.0055}{{\ttfamily 1311.0055}}].

\bibitem{Bechtle:2020pkv}
P.~Bechtle, D.~Dercks, S.~Heinemeyer, T.~Klingl, T.~Stefaniak, G.~Weiglein
  et~al., \emph{{HiggsBounds-5: Testing Higgs Sectors in the LHC 13 TeV Era}},
  \href{https://doi.org/10.1140/epjc/s10052-020-08557-9}{\emph{Eur. Phys. J. C}
  {\bfseries 80} (2020) 1211}
  [\href{https://arxiv.org/abs/2006.06007}{{\ttfamily 2006.06007}}].

\bibitem{Bechtle_2014}
P.~Bechtle, S.~Heinemeyer, O.~Stål, T.~Stefaniak and G.~Weiglein,
  \emph{Higgssignals: Confronting arbitrary higgs sectors with measurements at
  the tevatron and the lhc},
  \href{https://doi.org/10.1140/epjc/s10052-013-2711-4}{\emph{The European
  Physical Journal C} {\bfseries 74} (2014) }.

\bibitem{Bechtle:2020uwn}
P.~Bechtle, S.~Heinemeyer, T.~Klingl, T.~Stefaniak, G.~Weiglein and
  J.~Wittbrodt, \emph{{HiggsSignals-2: Probing new physics with precision Higgs
  measurements in the LHC 13 TeV era}},
  \href{https://doi.org/10.1140/epjc/s10052-021-08942-y}{\emph{Eur. Phys. J. C}
  {\bfseries 81} (2021) 145}
  [\href{https://arxiv.org/abs/2012.09197}{{\ttfamily 2012.09197}}].

\bibitem{ATLAS:2015kpj}
{\scshape ATLAS} collaboration, \emph{{Search for a CP-odd Higgs boson decaying
  to Zh in pp collisions at $\sqrt{s} = 8$ TeV with the ATLAS detector}},
  \href{https://doi.org/10.1016/j.physletb.2015.03.054}{\emph{Phys. Lett. B}
  {\bfseries 744} (2015) 163}
  [\href{https://arxiv.org/abs/1502.04478}{{\ttfamily 1502.04478}}].

\bibitem{CMS:2018xvc}
{\scshape CMS} collaboration, \emph{{Search for a heavy pseudoscalar boson
  decaying to a Z boson and a Higgs boson at sqrt(s)=13 TeV}}, .

\bibitem{ATLAS:2018oht}
{\scshape ATLAS} collaboration, \emph{{Search for a heavy Higgs boson decaying
  into a $Z$ boson and another heavy Higgs boson in the $\ell\ell bb$ final
  state in $pp$ collisions at $\sqrt{s}=13$ TeV with the ATLAS detector}},
  \href{https://doi.org/10.1016/j.physletb.2018.07.006}{\emph{Phys. Lett. B}
  {\bfseries 783} (2018) 392}
  [\href{https://arxiv.org/abs/1804.01126}{{\ttfamily 1804.01126}}].

\bibitem{CMS:2015mca}
{\scshape CMS} collaboration, \emph{{Search for additional neutral Higgs bosons
  decaying to a pair of tau leptons in $pp$ collisions at $\sqrt{s}$ = 7 and 8
  TeV}}, .

\bibitem{CMS:2017vpy}
{\scshape CMS} collaboration, \emph{{Search for a new scalar resonance decaying
  to a pair of Z bosons in proton-proton collisions at $\sqrt s$ = 13 TeV}}, .

\bibitem{ATLAS:2014jdv}
{\scshape ATLAS} collaboration, \emph{{Search for Scalar Diphoton Resonances in
  the Mass Range $65-600$ GeV with the ATLAS Detector in $pp$ Collision Data at
  $\sqrt{s}$ = 8 $TeV$}},
  \href{https://doi.org/10.1103/PhysRevLett.113.171801}{\emph{Phys. Rev. Lett.}
  {\bfseries 113} (2014) 171801}
  [\href{https://arxiv.org/abs/1407.6583}{{\ttfamily 1407.6583}}].

\bibitem{CMS:2013wyb}
{\scshape CMS} collaboration, \emph{{Properties of the Higgs-like boson in the
  decay H to ZZ to 4l in pp collisions at sqrt s =7 and 8 TeV}}, .

\bibitem{HFLAV:2016hnz}
{\scshape HFLAV} collaboration, \emph{{Averages of $b$-hadron, $c$-hadron, and
  $\tau$-lepton properties as of summer 2016}},
  \href{https://doi.org/10.1140/epjc/s10052-017-5058-4}{\emph{Eur. Phys. J. C}
  {\bfseries 77} (2017) 895}
  [\href{https://arxiv.org/abs/1612.07233}{{\ttfamily 1612.07233}}].

\bibitem{Misiak:2017bgg}
M.~Misiak and M.~Steinhauser, \emph{{Weak radiative decays of the B meson and
  bounds on $M_{H^\pm }$ in the Two-Higgs-Doublet Model}},
  \href{https://doi.org/10.1140/epjc/s10052-017-4776-y}{\emph{Eur. Phys. J. C}
  {\bfseries 77} (2017) 201}
  [\href{https://arxiv.org/abs/1702.04571}{{\ttfamily 1702.04571}}].

\bibitem{Mahmoudi_2009}
F.~Mahmoudi, \emph{Superiso v2.3: A program for calculating flavor physics
  observables in supersymmetry},
  \href{https://doi.org/10.1016/j.cpc.2009.02.017}{\emph{Computer Physics
  Communications} {\bfseries 180} (2009) 1579–1613}.

\bibitem{Alloul:2013bka}
A.~Alloul, N.~D. Christensen, C.~Degrande, C.~Duhr and B.~Fuks,
  \emph{{FeynRules 2.0 - A complete toolbox for tree-level phenomenology}},
  \href{https://doi.org/10.1016/j.cpc.2014.04.012}{\emph{Comput. Phys. Commun.}
  {\bfseries 185} (2014) 2250}
  [\href{https://arxiv.org/abs/1310.1921}{{\ttfamily 1310.1921}}].

\bibitem{Alwall:2011uj}
J.~Alwall, M.~Herquet, F.~Maltoni, O.~Mattelaer and T.~Stelzer, \emph{{MadGraph
  5 : Going Beyond}},
  \href{https://doi.org/10.1007/JHEP06(2011)128}{\emph{JHEP} {\bfseries 06}
  (2011) 128} [\href{https://arxiv.org/abs/1106.0522}{{\ttfamily 1106.0522}}].

\bibitem{Alwall:2014hca}
J.~Alwall, R.~Frederix, S.~Frixione, V.~Hirschi, F.~Maltoni, O.~Mattelaer
  et~al., \emph{{The automated computation of tree-level and next-to-leading
  order differential cross sections, and their matching to parton shower
  simulations}}, \href{https://doi.org/10.1007/JHEP07(2014)079}{\emph{JHEP}
  {\bfseries 07} (2014) 079} [\href{https://arxiv.org/abs/1405.0301}{{\ttfamily
  1405.0301}}].

\bibitem{Sjostrand:2006za}
T.~Sjostrand, S.~Mrenna and P.~Z. Skands, \emph{{PYTHIA 6.4 Physics and
  Manual}}, \href{https://doi.org/10.1088/1126-6708/2006/05/026}{\emph{JHEP}
  {\bfseries 05} (2006) 026}
  [\href{https://arxiv.org/abs/hep-ph/0603175}{{\ttfamily hep-ph/0603175}}].

\bibitem{Sjostrand:2014zea}
T.~Sjöstrand, S.~Ask, J.~R. Christiansen, R.~Corke, N.~Desai, P.~Ilten et~al.,
  \emph{{An Introduction to PYTHIA 8.2}},
  \href{https://doi.org/10.1016/j.cpc.2015.01.024}{\emph{Comput. Phys. Commun.}
  {\bfseries 191} (2015) 159}
  [\href{https://arxiv.org/abs/1410.3012}{{\ttfamily 1410.3012}}].

\bibitem{deFavereau:2013fsa}
{\scshape DELPHES 3} collaboration, \emph{{DELPHES 3, A modular framework for
  fast simulation of a generic collider experiment}},
  \href{https://doi.org/10.1007/JHEP02(2014)057}{\emph{JHEP} {\bfseries 02}
  (2014) 057} [\href{https://arxiv.org/abs/1307.6346}{{\ttfamily 1307.6346}}].

\bibitem{Cacciari:2008gp}
M.~Cacciari, G.~P. Salam and G.~Soyez, \emph{{The Anti-k(t) jet clustering
  algorithm}}, \href{https://doi.org/10.1088/1126-6708/2008/04/063}{\emph{JHEP}
  {\bfseries 04} (2008) 063} [\href{https://arxiv.org/abs/0802.1189}{{\ttfamily
  0802.1189}}].

\bibitem{LHCHiggsCrossSectionWorkingGroup:2016ypw}
{\scshape LHC Higgs Cross Section Working Group} collaboration, \emph{{Handbook
  of LHC Higgs Cross Sections: 4. Deciphering the Nature of the Higgs Sector}},
   \href{https://arxiv.org/abs/1610.07922}{{\ttfamily 1610.07922}}.

\bibitem{Campanario:2010hp}
F.~Campanario, C.~Englert, S.~Kallweit, M.~Spannowsky and D.~Zeppenfeld,
  \emph{{NLO QCD corrections to WZ+jet production with leptonic decays}},
  \href{https://doi.org/10.1007/JHEP07(2010)076}{\emph{JHEP} {\bfseries 07}
  (2010) 076} [\href{https://arxiv.org/abs/1006.0390}{{\ttfamily 1006.0390}}].

\bibitem{Cascioli:2014yka}
F.~Cascioli, T.~Gehrmann, M.~Grazzini, S.~Kallweit, P.~Maierhöfer, A.~von
  Manteuffel et~al., \emph{{ZZ production at hadron colliders in NNLO QCD}},
  \href{https://doi.org/10.1016/j.physletb.2014.06.056}{\emph{Phys. Lett.}
  {\bfseries B735} (2014) 311}
  [\href{https://arxiv.org/abs/1405.2219}{{\ttfamily 1405.2219}}].

\bibitem{Maltoni:2015ena}
F.~Maltoni, D.~Pagani and I.~Tsinikos, \emph{{Associated production of a
  top-quark pair with vector bosons at NLO in QCD: impact on $
  \mathrm{t}\overline{\mathrm{t}}\mathrm{H} $ searches at the LHC}},
  \href{https://doi.org/10.1007/JHEP02(2016)113}{\emph{JHEP} {\bfseries 02}
  (2016) 113} [\href{https://arxiv.org/abs/1507.05640}{{\ttfamily
  1507.05640}}].

\bibitem{Artoisenet_2013}
P.~Artoisenet, R.~Frederix, O.~Mattelaer and R.~Rietkerk, \emph{Automatic
  spin-entangled decays of heavy resonances in monte carlo simulations},
  \href{https://doi.org/10.1007/jhep03(2013)015}{\emph{Journal of High Energy
  Physics} {\bfseries 2013} (2013) }.

\bibitem{Cowan:2010js}
G.~Cowan, K.~Cranmer, E.~Gross and O.~Vitells, \emph{{Asymptotic formulae for
  likelihood-based tests of new physics}},
  \href{https://doi.org/10.1140/epjc/s10052-011-1554-0}{\emph{Eur. Phys. J. C}
  {\bfseries 71} (2011) 1554}
  [\href{https://arxiv.org/abs/1007.1727}{{\ttfamily 1007.1727}}].

\end{thebibliography}\endgroup
%
%
%
%
%
%
\end{document}